G Sun et al.

# Investigating the influence of divertor baffles on nitrogen-seeded detachment in TCV with SOLPS-ITER simulations and TCV experiments

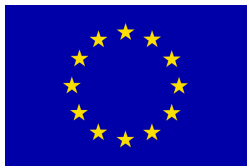

This work has been carried out within the framework of the EUROfusion Consortium and has received funding from the Euratom research and training programme 2014-2018 and 2019-2020 under grant agreement No 633053. The views and opinions expressed herein do not necessarily reflect those of the European Commission.

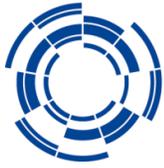

Editor
0 - Please choose a journal

PROGRAMME MANAGEMENT UNIT
**Dr. Kinga GÁL**
Scientific Advisor
Phone: +49 89 3299 1966
Kinga.Gal@euro-fusion.org
Secretariat: Tel: +49 89 3299 4201

September 19, 2024

**Subject:** Paper entitled *"Investigating the influence of divertor baffles on nitrogen-seeded detachment in TCV with SOLPS-ITER simulations and TCV experiments"* by G Sun

Dear Editor of 0 - Please choose a journal,

Please find enclosed the manuscript: *Investigating the influence of divertor baffles on nitrogen-seeded detachment in TCV with SOLPS-ITER simulations and TCV experiments* by G Sun for consideration of publication in 0 - Please choose a journal.

I would appreciate if you would send the correspondence regarding the refereeing process to the lead author: guang-yu.sun@epfl.ch.

With best regards,

Dr. Kinga GÁL
EUROfusion Scientific Advisor

**EUROfusion**
Programme Management Unit
Boltzmannstr. 2
85748 Garching
Germany

info@euro-fusion.org
www.euro-fusion.org

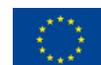

EUROfusion receives funding
from the European Commission
under Grant Agreement
No. 101052200

# Investigating the influence of divertor baffles on nitrogen-seeded detachment in TCV with SOLPS-ITER simulations and TCV experiments

G. Sun, [1] H. Reimerdes, C. Theiler, B.P. Duval, M. Carpita, C. Colandrea, R. Ducker, O. Février, S. Gorno, L. Simons, E. Tonello, the EUROfusion tokamak exploitation team[2] and the TCV team[3]

*Ecole Polytechnique Fédérale de Lausanne (EPFL), Swiss Plasma Center (SPC), CH-1015 Lausanne, Switzerland*

Plasma edge simulations with the SOLPS-ITER code are performed to study the influence of divertor baffles on nitrogen-seeded detachment in TCV single-null, L-mode discharges. Scans of nitrogen seeding rate are conducted in both baffled and unbaffled TCV divertors, where the nitrogen seeding with baffles is found to yield lower target temperatures and heat fluxes than with baffles-only and with seeding-only. The cumulative effects of baffles and seeding on target parameters are explained by the two-point model. The divertor neutral density and neutral compression increase with baffles, due to lower divertor to main chamber neutral conductance, as explained by a schematic neutral transport model with baffles. The nitrogen retention, defined as the ratio of average nitrogen nuclei density in divertor and main chamber, increases with the seeding rate if baffled, and remains constant if unbaffled. At the same outboard mid-plane separatrix plasma density, the nitrogen retention with baffles is lower than the unbaffled retention at low seeding levels and is higher at high seeding levels, which is explained by the changes of nitrogen ion and neutral transport with baffles and seeding. The baffled carbon retention is higher than the unbaffled retention due to lower divertor to main chamber carbon neutral conductance. Baffles increase the divertor radiation. The predicted trends of target parameters, the distribution of neutrals and radiations are well supported by TCV experiments, though discrepancies in the absolute values remain. The simulations yield an overall colder and denser divertor, consistent with previous SOLPS-ITER simulations of Ohmically heated L-modes in TCV. The successful comparison of simulation and experiment, together with the understanding gained from the neutral transport model, increases the confidence in the SOLPS simulations for the next TCV divertor upgrade.

## 1. INTRODUCTION

The divertor targets of a fusion reactor based on the tokamak concept will be exposed to intense plasma heat fluxes. Controlling plasma-surface interactions at the targets within sustainable levels will likely require operating its divertor in the detached regime, characterized by low target plasma temperatures and reduced particle and heat fluxes [1-3]. Accessing a detached divertor requires sufficient power dissipation in the divertor volume, which is commonly achieved by operating at high plasma density and/or by seeding impurities [4-7]. Increasing divertor closure can further facilitate the access to detachment [8-11]. The tokamak à configuration variable (TCV) [12] contributes to the development and optimization of a divertor solution for a fusion reactor through proof-of-principle experiments and plasma edge model validation.

---

[1] Author to whom correspondence should be addressed: guang-yu.sun@epfl.ch
[2] See the author list of E. Joffrin et al. Nuclear Fusion 64 112019 (2024)
[3] See the author list of H. Reimerdes et al. Nuclear Fusion 62 042018 (2022)

Impurity seeding converts a portion of the scrape-off layer (SOL) heat flux into radiation, which distributes power more evenly across the vessel wall, rather than locally concentrating it onto the divertor targets. Nitrogen and noble gases (e.g. Ne, Ar, Kr) are commonly used impurity species. However, high core impurity concentration can lead to excessive confinement degradation or intolerable fuel dilution [13]. Impurities with higher charge numbers, Z, have higher radiation efficiency in the plasma core and are more prone to lead to excessive core radiation or even a discharge collapse [14, 15]. Low-Z impurities radiate less in the plasma core but dilute the fusion fuel, which may be intolerable in a reactor [4]. Since the radiation efficiencies of impurities depends on the electron temperature [4, 16], the optimum choice of impurity species is device dependent. Nitrogen is particularly suited for medium size devices, and has been widely used in TCV experiments [17-19].

Increasing divertor closure was found to increase the confinement of neutral particles in the divertor and enhance the transfer of momentum and energy from the plasma to neutrals [20-22]. Flexible baffles have been installed in TCV during dedicated campaigns to explore the role of neutrals for plasma exhaust [23]. Too much divertor closure can, however, lead to excessive heat fluxes and recycling on the main-plasma-facing side of the baffles [24]. The optimal location and closure of TCV baffles have been investigated in detail through both simulations and experiments [24-26].

Previous TCV simulations investigated neutral baffling without impurity seeding [27] and seeding without baffles [28]. The present work extends the investigations to nitrogen-seeded detachment with baffles, combining SOLPS-ITER simulations and TCV experiments, to provide an improved understanding of the effect of baffles on deuterium and impurity transport. Impurities are subject to thermal and friction forces, with the parallel force balance and E×B drifts determining the impurity leakage from the divertor into the main chamber. Recent work suggests that the relative location between the impurity stagnation point, where the impurity flow velocity near the separatrix is zero, and the impurity ionization front regulates the impurity leakage [29]. Baffles affect both locations and inevitably complicate the behavior of impurities, motivating the investigations of the interplay between baffling and seeding.

The paper is organized as follows. Section 2 introduces the adopted SOLPS-ITER simulation model. In Section 3, simulated upstream and target profiles, neutral compression, impurity and radiation distributions are analyzed in both baffled and unbaffled conditions. A schematic neutral transport model is introduced to explain the effect of baffles on the neutral deuterium and impurity distribution. In Section 4, the model predictions are compared with TCV experiments. Trends identified in Section 3 are validated and the discrepancies are discussed. Concluding remarks are given in Section 5.

2. SIMULATION MODEL

This section introduces the SOLPS-ITER simulation setup. SOLPS-ITER is a 2D transport code to simulate the plasma edge of magnetic fusion devices [30, 31]. It combines the 2D multi-fluid plasma transport model B2.5 and the Monte-Carlo neutral transport kinetic model EIRENE [32], and is used to predict the divertor performance in ITER and in various reactor designs. The B2.5 model solves for the plasma parameters (density $n$, parallel ion velocity $v_{i,//}$, temperature $T$, potential $\varphi$, etc.) from the balance equations of particle number, parallel momentum, energy and charge. Meanwhile, EIRENE simulates the plasma-neutral interactions and calculates the source terms in the plasma fluid equations in B2.5. The surface reflections rely on the TRIM database, the radiation is evaluated using ADAS, and reaction rates (cf. Table I) are obtained from AMJUEL, HYDHEL and METHANE. The currently adopted code version is 3.0.8.

The considered equilibrium is a TCV lower single null configuration with a plasma current of 250 kA and a magnetic field of 1.4 T, typically used in Ohmically heated L-mode density ramp and seeding experiments, Figure 1(a). The generated curvilinear quadrangular B2.5 computational grid, Figure 1(b), covers the scrape-off layer (SOL) up to the baffles, whereas the triangular EIRENE mesh, Figure 1(c), extends to the vessel wall. The geometry of the short inner and long outer baffles (referred to as the SI-LO configuration) corresponds to the experiments and is the same as in previous simulations [27]. All plasma-facing components are assumed to consist of carbon. Since the plasma grid is limited by the baffle tips, the sheath boundary condition can only be applied at the target plates, whereas constant density and temperature fall-off lengths are set for far SOL boundaries. Radial plasma flow across the far SOL plasma grid boundary is returned as neutrals, which may result in an overestimation of the recycling on the main chamber vessel wall. The problem is avoided in other edge codes such as SolEdge2D [33] and efforts are underway to extend also the plasma grid in SOLPS-ITER to the wall [34].

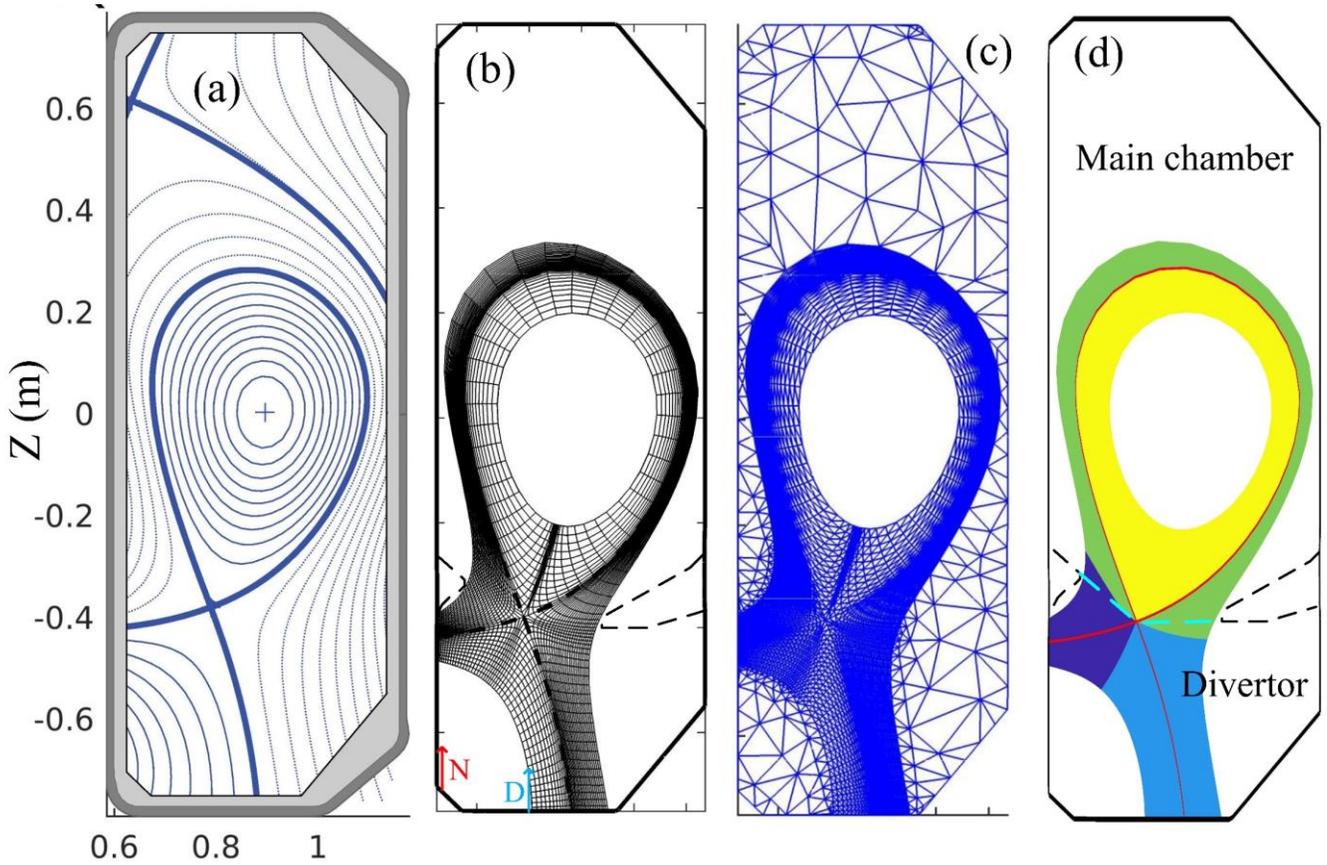

Figure 1. Employed magnetic equilibrium and generated simulation grids. (a) Equilibrium of TCV discharge 68313 at t=1.0 s. (b) B2.5 grid with nitrogen and deuterium puff locations marked with arrows and optional baffles with dashed lines. (c) EIRENE grid for the unbaffled option. (d) Distinct regions of the B2.5 grid, including inner divertor (dark blue), outer divertor (light blue), upper SOL (green) and core (yellow). Main chamber and divertor regions are separated by line segments (cyan) connecting the X-point and the baffle tips.

Considered plasma species include all charge states of deuterium ($D_2, D, D^+, D_2^+$), carbon ($C - C^{6+}$) and nitrogen ($N - N^{7+}$). Due to the short disassociation mean free path of $N_2$, only atomic $N$ is considered. In addition, chemical reactions such as the formation of nitrogen hydrides and hydrocarbon are neglected. Detailed reactions are listed in Table I. Note, that recent studies highlighted a greater importance of plasma-molecule interaction, with systematic errors in some of the code-embedded reaction cross sections [35]. For the present work, the default reaction list is used, without the correction for plasma-molecule interaction. Deuterium and nitrogen are puffed in the divertor private flux region (PFR), Figure 1(b), as typically done in TCV experiments. Ions and neutral atoms colliding with the target are either recycled as thermalized molecules or reflected as fast atoms, as described by EIRENE and TRIM [36]. The recycling coefficient, defined as the ratio of recycled neutral flux and incident particle flux $R = \Gamma^{recycled}/\Gamma^{in}$, is set to 0.99 for deuterium and 0 for carbon. With a previous TCV study showing that a recycling coefficient of 1.0 for neutral nitrogen and 0.3-0.5 for nitrogen ions provide the best match between the spectroscopic measurements of nitrogen emission lines and SOLPS simulation results [28], a value of 0.3 is chosen. Ion and neutral particle impact onto the wall also release carbon through physical and/or chemical sputtering. The former is calculated with the Roth-Bogdansky formula [37], whereas the latter has no practical energy threshold and a value of 3.5% is chosen following previous SOLPS simulation for TCV [27]. The selected chemical sputtering yield is comparable with the TCV experiment results, and produces edge carbon concentration consistent with TCV radiation measurements [38].

TABLE I. Reactions considered in the simulations and corresponding database.

| Index | Reference code | Type | Reaction |
|---|---|---|---|
| | | AMJUEL | |
| 1 | H.4/H.10 2.1.5 | Ionization | $D + e \rightarrow D^+ + 2e$ |
| 2 | H.4/H.10 2.6A0 | Ionization | $C + e \rightarrow C^+ + 2e$ |
| 3 | H.4/H.10 2.7A0 | Ionization | $N + e \rightarrow N^+ + 2e$ |
| 4 | H4 2.2.9 | Ionization | $D_2 + e \rightarrow D_2^+ + 2e$ |
| 5 | H4 2.2.5g | Dissociation | $D_2 + e \rightarrow D + D + e$ |
| 6 | H4 2.2.10 | Ionization/dissociation | $D_2 + e \rightarrow D^+ + D + 2e$ |
| 7 | H.0/H.1/H.3 0.3T | Elastic collision | $D_2 + D^+ \rightarrow D_2 + D^+$ |
| 8 | H2 3.2.3 | Charge exchange | $D_2 + D^+ \rightarrow D_2^+ + D$ |
| 9 | H4 2.2.12 | Dissociation | $D_2^+ + e \rightarrow D^+ + D + e$ |
| 10 | H4 2.2.11 | Ionization/dissociation | $D_2^+ + e \rightarrow D^+ + D^+ + 2e$ |
| 11 | H.4/H.8 2.2.14 | Dissociation/recombination | $D_2^+ + e \rightarrow D + D$ |
| 12 | H.4/H.10 2.1.8 | Recombination | $D^+ + e \rightarrow D$ |
| | | HYDHEL | |
| 13 | H.1/H.3 3.1.8 | Charge exchange | $D^+ + D \rightarrow D + D^+$ |
| | | AMMONX | |
| 14 | H.2 R-H-H | Association/ionization | $D + D \rightarrow D_2^+ + e$ |
| 15 | H.2 R-H-H2 | Association/ionization | $D + D_2 \rightarrow D_3^+ + e$ |
| 16 | H.2 R-H2-H | Association | $D + D_2 \rightarrow 3D$ |
| 17 | H.2 R-H2-H2 | Dissociation | $D_2 + D_2 \rightarrow D_2 + D + D$ |
| | | METHAN | |
| 18 | H.1/H.3 3.2 | Charge exchange | $D^+ + C \rightarrow D + C^+$ |
| | | ADAS | |
| 19 | C H.4 acd96/ H.10prb96 | Recombination | $C^+ + e \rightarrow C + h\nu$ |
| 20 | N H.4 acd96/ H.10prb96 | Recombination | $N^+ + e \rightarrow N + h\nu$ |

Anomalous cross field transport is approximated by constant empirical transport coefficients, with $D_\perp = 0.2\ m^2 s^{-1}$ and $\chi_{\perp,e} = \chi_{\perp,i} = 1.0\ m^2 s^{-1}$ as in previous simulations [27, 28], providing reasonable agreement with upstream profiles measured by Thomson scattering in Ohmically heated L-mode TCV discharges, Section 4. A constant input power of 330 kW equally shared between the electrons and ions is set as the core boundary condition. Since this work aims at a qualitative assessment of nitrogen seeding and neutral baffling, drifts are deactivated as they require smaller time steps and cause numerical instabilities. Drifts are known to cause asymmetrical particle and power distributions in the divertor [14, 39, 40].

## 3. SIMULATION OF NITROGEN SEEDING IN BAFFLED AND UNBAFFLED TCV CONFIGURATIONS

In this section, SOLPS simulations are used to analyze the influence of nitrogen seeding and neutral baffling on detachment. The discussion starts with the upstream and target profiles, and then covers the behavior of neutrals and impurities.

Discharge parameters are chosen from Ohmically heated L-mode experiments that explore the transition from the conduction-limited (high-recycling) divertor regime to the detached regime. This transition typically requires the target electron temperature, $T_{e,t}$, to decrease below 5 eV. A decrease of the peak target electron temperature below 5 eV is, therefore, used as a proxy for the onset of detachment. While the nitrogen seeding rate is varied up to $8 \times 10^{20}$ atom/s with and without baffles, the deuterium fueling rates are adjusted from $2.5 \times 10^{20}$ molecules/s to $1.5 \times 10^{21}$ molecules/s to maintain an outboard midplane separatrix density, $n_{e,sep} = 1.5 \times 10^{19}$ m$^{-3}$.

### 3.1. Upstream and target profiles

The change of upstream and target properties including upstream and outer target temperatures, densities, target particle and heat fluxes show the effects of baffles and nitrogen seeding on detachment. Two-point model (2PM) analyses is performed to interpret the simulation results. In TCV, the outer divertor has a longer connection length and usually detaches first. The following analysis, therefore, focuses on the outer divertor.

#### 3.1.1 Observations

At constant upstream (separatrix at the outer mid-plane) density, $n_{e,u} = 1.5 \times 10^{19}$ m$^{-3}$, the target electron temperature and heat flux (including thermal, kinetic, and potential energy fluxes) decrease with both, seeding and baffling, Figure 2(a, b). The target electron density, $n_{e,t}$, is higher with baffles, and is insensitive to the seeding rate, Figure 2(c). The total target particle flux of all ion species is also higher with baffles, but decreases with increasing seeding rate, Figure 2(d). The contribution of impurity particle flux in the total particle flux increases from 5% to 20% from zero to maximum seeding levels. The unbaffled outer divertor detaches (maximum $T_{e,t} < 5$ eV) for seeding rates above $5 \times 10^{20}$ atoms/s and the baffled divertor is already detached without seeding, Figure 2(a).

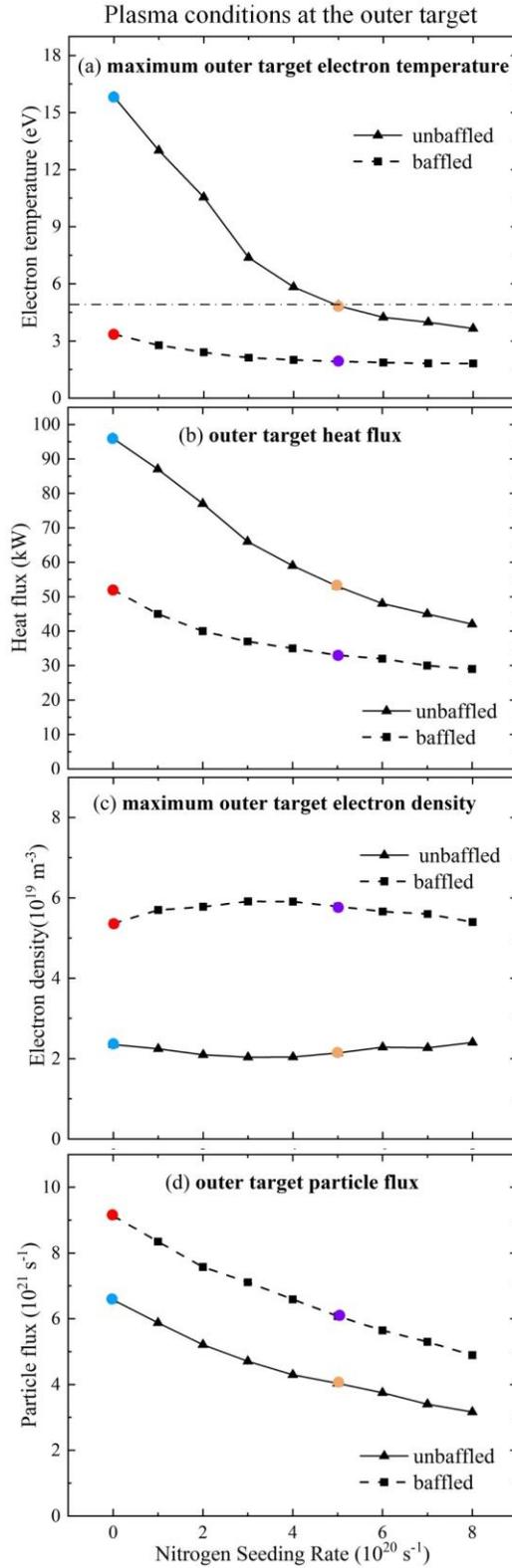

Figure 2. Dependence of (a) the outer target peak electron temperature, (b) parallel heat flux, (c) peak electron density, and (d) total parallel particle flux of all plasma species on nitrogen seeding rate at an upstream separatrix density, $n_{e,sep}= 1.5 \times 10^{19}$ m$^{-3}$. The colored points

correspond to the four simulation cases highlighted in Table II. The proxy for the detachment threshold is marked by a dash-dotted horizontal line in (a).

Four simulation cases from the seeding rate scans, Figure 2, are selected and compared to further illustrate the separate effects of nitrogen seeding and baffling, as well as the combination of both, on upstream and target parameters. They consist of an unbaffled, unseeded case (case 1), a baffled, unseeded case (case 2), an unbaffled, seeded case (case 3), and a baffled, seeded case (case 4), Table II. The unbaffled, unseeded divertor (case 1) is attached while all other cases are detached. A nitrogen seeding rate of $5 \times 10^{20}$ atoms/s is chosen such that the unbaffled, seeded divertor (case 3) and the baffled, unseeded divertor (case 2) have approximately the same outer target heat flux, Figure 2(b), and are, therefore, similarly favorable for power exhaust at the outer divertor target. This nitrogen seeding rate is comparable with rates used in TCV seeding experiments, and leads to a $Z_{eff}$ of 1.9 at the outboard midplane separatrix (without baffles), which is considered tolerable in a reactor scenario [41].

TABLE II. Key input and output parameters in four selected simulation cases. The color-coding of each case is kept throughout this work.

| Case | Type | 1 | 2 | 3 | 4 |
|---|---|---|---|---|---|
| Baffles | | No | Yes | No | Yes |
| N seeding rate ($10^{20}$ s$^{-1}$) | Input | 0.0 | 0.0 | 5.0 | 5.0 |
| D$_2$ fueling rate ($10^{20}$ s$^{-1}$) | Input | 2.9 | 7.3 | 2.5 | 6.7 |
| Upstream electron separatrix temperature (eV) | Output | 50.6 | 48.1 | 52.4 | 47.6 |
| Upstream electron separatrix density ($10^{19}$ m$^{-3}$) | Output | 1.48 | 1.51 | 1.49 | 1.52 |
| Outer target maximum electron temperature (eV) | Output | 15.8 | 3.34 | 4.83 | 1.92 |
| Outer target maximum electron density ($10^{19}$ m$^{-3}$) | Output | 2.35 | 5.64 | 2.14 | 5.78 |
| Outer target heat flux (kW) | Output | 96 | 52 | 53 | 33 |
| Inner target heat flux (kW) | Output | 140 | 138 | 90 | 82 |

Baffles increase the neutral confinement in the divertor and reduce the neutral density in the main chamber, leading to a lower ionization rate inside the last-closed-flux-surface (LCFS), Figure 3(a), and, therefore, a lower cross-field particle flux. Since the simulations assume a constant cross-field diffusivity, the electron density gradient inside the LCFS decreases. The corresponding electron temperature gradient increases accordingly due to the constant power entering from the core boundary in these simulations, Figure 3(b). Meanwhile, nitrogen seeding barely affects the gradients inside the LCFS, consistent with previous SOLPS simulations of unbaffled impurity seeding [28, 42, 43]. Neither the baffles nor seeding have a significant effect on the upstream SOL profiles at constant upstream separatrix density. As a consequence, the electron pressure in the upstream SOL also remains unaffected by the simulated baffles and/or seeding rates.

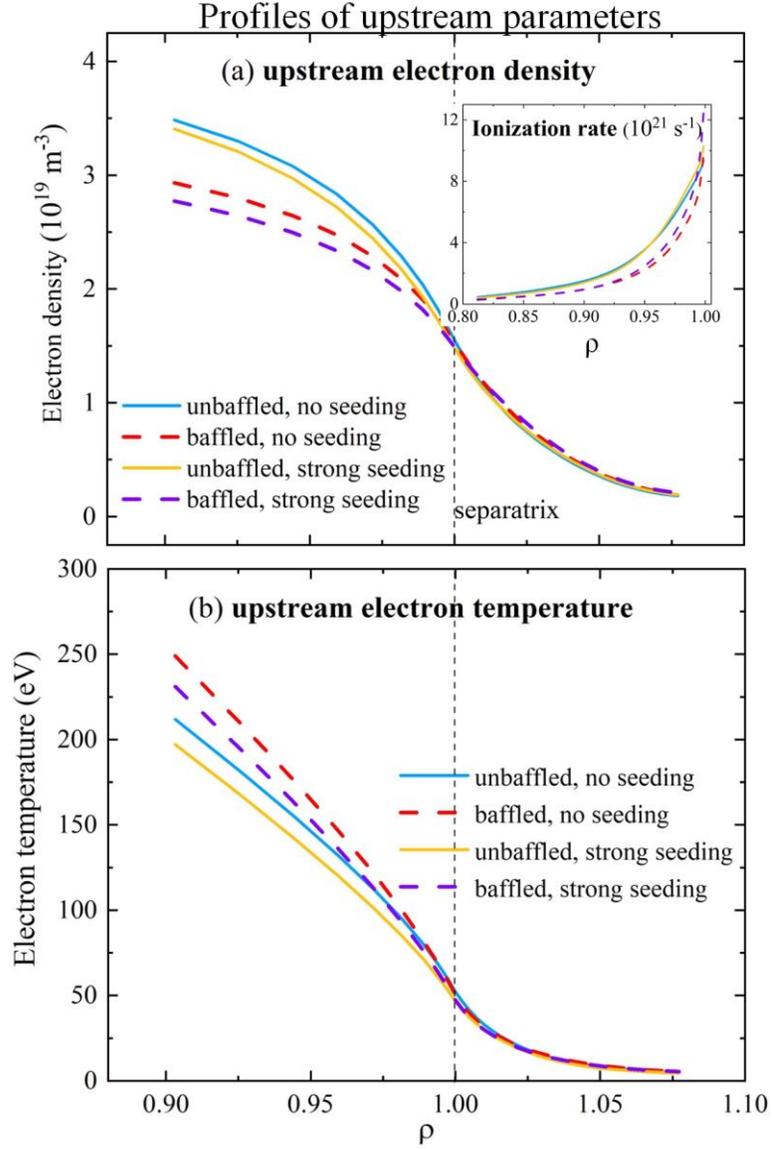

Figure 3. Upstream (a) electron density and ionization rate (insert) and (b) electron temperature profiles of the four simulation cases listed in Table II. The upstream separatrix electron density is fixed at $1.5 \times 10^{19}$ m$^{-3}$ and the nitrogen seeding rate is $5.0 \times 10^{20}$ s$^{-1}$ in the seeded cases.

To understand the effect of baffles on the access to divertor detachment, the target parameter profiles are analyzed. Baffles increase $n_{e,t}$ and decrease $T_{e,t}$ across the entire target, whereas impurity seeding leaves $n_{e,t}$ almost unchanged and primarily decreases $T_{e,t}$, Figure 4(a, b). The Bohm criterion at the target sheath entrance, employed as boundary condition at the divertor targets in these simulations, links $n_{e,t}$, $T_{e,t}$ and the target parallel particle flux density along magnetic field lines:

$$\Gamma_t = n_{e,t} c_{s,t} \tag{1}$$

where the sound speed $c_{s,t} = \sqrt{(T_{e,t} + T_{i,t})/m_D}$ depends on the target electron and ion temperatures, with $m_D$ being the main ion mass. The target parallel heat flux density along magnetic field lines including plasma thermal and kinetic energy and potential energy of ionization is approximated by:

$$q_t^{TKP} = \gamma_e T_{e,t} \Gamma_t + \gamma_i T_{i,t} \Gamma_t + \varepsilon_{pot} \Gamma_t = q_t^{TK} + \varepsilon_{pot} \Gamma_t \tag{2}$$

where $\gamma_e$ and $\gamma_i$ are the electron and ion sheath heat transmission coefficients and $\varepsilon_{pot}$ is the surface recombination energy. The superscript 'TKP' refers to the considered thermal, kinetic and potential energy contribution and the superscript "TK" indicates that the potential energy contribution is excluded. Note that heat flux contributions from radiation and neutrals are not considered. The increase of $n_{e,t}$ with baffles is sufficiently large to counteract the decrease of $T_{e,t}$, resulting in a larger particle flux with baffles, Figure 4(c). Baffles result in a lower heat flux, Figure 4(d), since $q_t^{TKP}$ has stronger dependence on $T_{e,t}$ than the particle flux, Equation (2). Meanwhile, seeding decreases $T_{e,t}$ whilst not significantly affecting $n_{e,t}$, resulting in a lower particle and heat flux. A combination of baffling and seeding results in the lowest target temperatures, Figure 4(a), the lowest target heat flux, Figure 4(d), and an intermediate particle flux, Figure 4(c), consistent with the cumulative effects of baffling and seeding. Though baffling-only (case 2) and seeding-only (case 3) have similar $q_t$ profile, the former features a larger $T_{e,t}$ reduction while the latter achieves heat flux reduction by decreasing $\Gamma_t$ in addition to a lesser $T_{e,t}$ decrease.

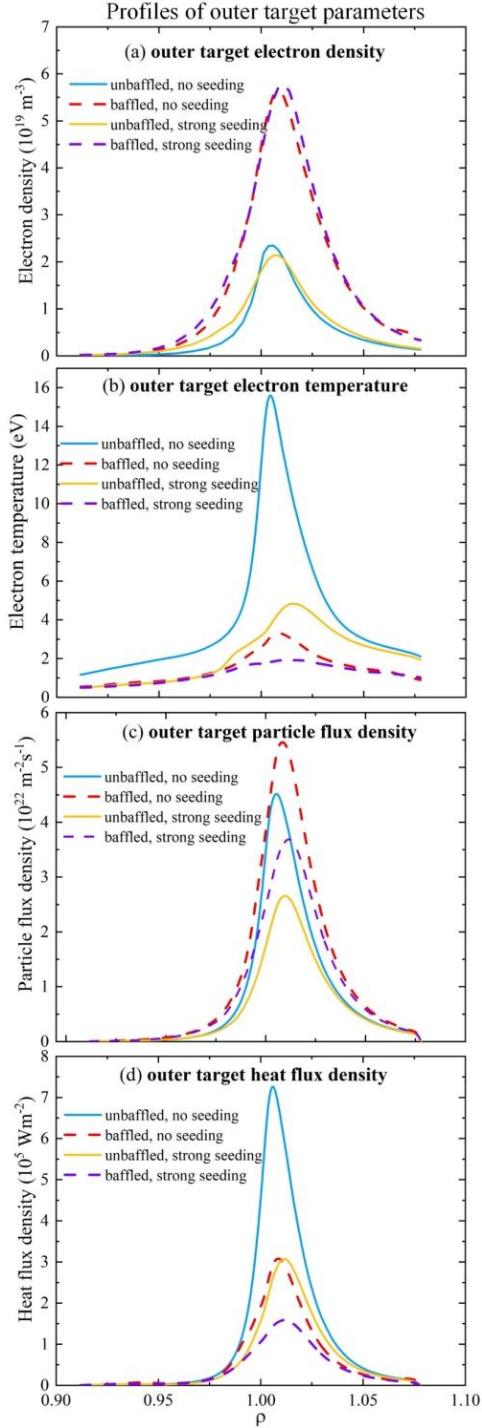

Figure 4. Outer target (a) electron density, (b) electron temperature, (c) parallel particle flux density of deuterium and (d) parallel heat flux density ($q_t^{TKP}$) profiles of the four simulation cases in Table II. The upstream separatrix electron density is fixed at $1.5 \times 10^{19}$ m$^{-3}$ and the nitrogen seeding rate is $5.0 \times 10^{20}$ s$^{-1}$ in the seeded cases.

3.1.2. Two-point model analyses

The changes of target parameters with baffles and seeding are interpreted through two-point model. According to the modified 2PM model, the target density, temperature and parallel particle flux density are [1]:

$$T_{e,t}^{2PM} = \frac{8m_D}{e\gamma^2} \frac{q_u^{TK2}}{p_{u,tot}^2} \left(\frac{R_u}{R_t}\right)^2 \frac{(1-f_{cooling})^2}{(1-f_{mom,loss})^2} \tag{3}$$

$$n_{e,t}^{2PM} = \frac{\gamma^2}{32m_D} \frac{p_{u,tot}^3}{q_u^{TK2}} \left(\frac{R_t}{R_u}\right)^2 \frac{(1-f_{mom,loss})^3}{(1-f_{cooling})^2} \tag{4}$$

$$\Gamma_t^{2PM} = \frac{\gamma}{8m_D} \frac{p_{u,tot}^2}{q_u^{TK}} \left(\frac{R_t}{R_u}\right) \frac{(1-f_{mom,loss})^2}{(1-f_{cooling})^1} \tag{5}$$

Here $\gamma = \gamma_e + \gamma_i T_i/T_e$ is the sheath heat transmission coefficient, $q_u^{TK}$ is the upstream parallel heat flux considering thermal and kinetic energy contribution, $p_{u,tot}$ is the total upstream pressure, and $R_t/R_u$ is the change in radius from upstream to target. The two volumetric loss terms $f_{mom,loss}$ and $f_{cooling}$ represent the fraction of momentum and power loss, defined as:

$$f_{mom,loss} \equiv 1 - \frac{p_{t,tot}}{p_{u,tot}} \tag{6}$$

$$f_{cooling} \equiv 1 - \frac{q_t^{TK} R_t}{q_u^{TK} R_u} \tag{7}$$

Here $p_{t,tot}$ and $q_t^{TK}$ are the target total pressure and parallel heat flux density considering thermal and kinetic energy contribution, respectively. Since $q_u^{TK}$ and $p_{u,tot}$ are similar in all four cases, baffles and seeding primarily affect target parameters by changing the two loss fraction terms. Note that the decrease of target thermal and kinetic heat flux depends solely on the increase of $f_{cooling}$, whereas target temperature, density and particle flux are affected by both, $f_{mom,loss}$ and $f_{cooling}$.

The volumetric loss terms are calculated from the simulations and used to interpret the target parameters. Baffles and nitrogen seeding, both, increase $f_{cooling}$, Figure 5(a). Note that the greater increase of $f_{cooling}$ with baffling (case 2) than with seeding (case 3) is a consequence of the choice of the seeding rate in case 3 to approximately match $q_t^{TKP}$ of case 2 (whereas the potential energy flux is not considered here). Baffles and seeding both increase $f_{mom,loss}$, similar to $f_{cooling}$, Figure 5(b). The increase of $f_{mom,loss}$ is due to higher momentum loss via charge exchange at lower plasma temperature. Note that the change of $f_{cooling}$ and $f_{mom,loss}$ with baffles in the far SOL are less important, where the values of target parameters are much lower than near the separatrix.

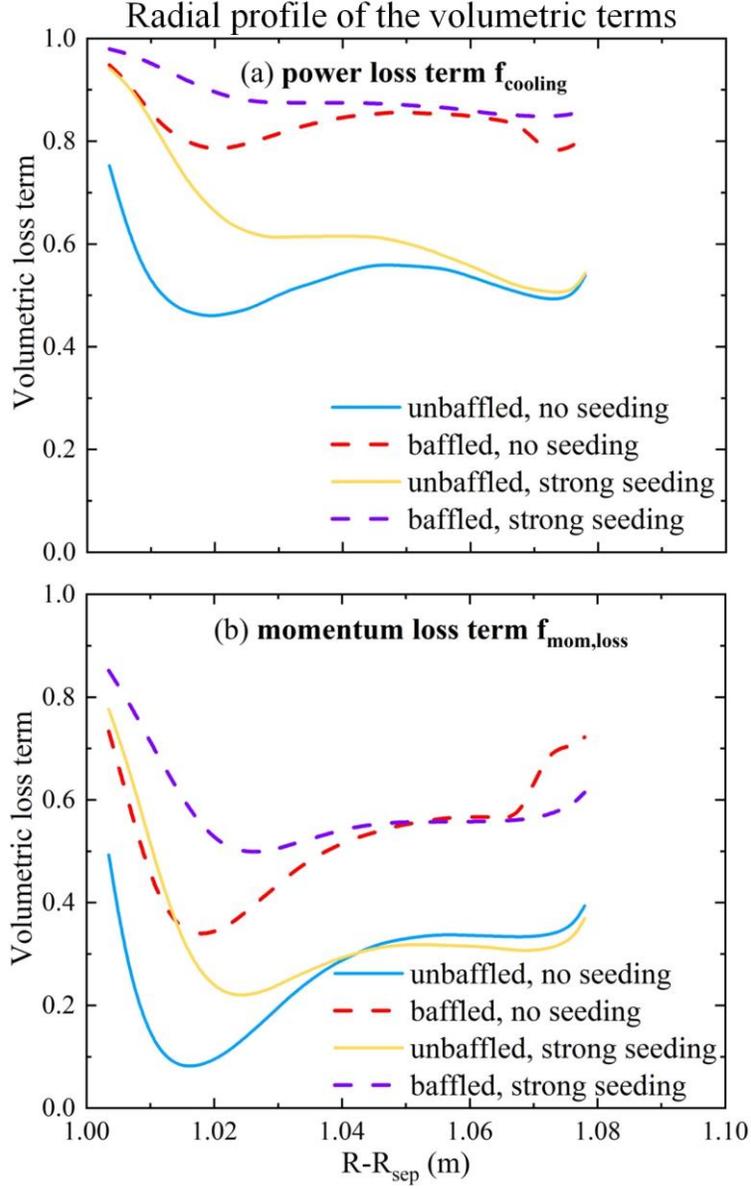

Figure 5. The volumetric power (a) and momentum (b) loss terms of the four cases in Table II. The upstream separatrix electron density is fixed at $1.5 \times 10^{19}$ m$^{-3}$ and the nitrogen seeding rate set to $5.0 \times 10^{20}$ s$^{-1}$ in the seeded cases.

Adding the baffles leads to a larger decrease of $1 - f_{cooling}$ than $1 - f_{mom,loss}$, which explains a lower $T_{et}$, and a higher $n_{et}$ and $\Gamma_t$. Increasing the seeding rate causes the factor $\frac{(1-f_{cooling})^2}{(1-f_{mom,loss})^2}$ to decrease, which explains a lower $T_{et}$. The factor $\frac{(1-f_{mom,loss})^3}{(1-f_{cooling})^2}$ varies little with the seeding rate, which is consistent with an $n_{et}$ that is insensitive to the seeding rate. The target particle flux $\Gamma_t \propto n_{et}\sqrt{T_{et}}$ consequently decreases with seeding.

## 3.2. Neutral transport

The volumetric losses, described by $f_{cooling}$ and $f_{mom,loss}$ in Section 3.1, are governed by the divertor neutral and impurity distributions through plasma-neutral interaction and impurity radiation. In this section, the effects of TCV gas baffles and nitrogen seeding on neutral properties are investigated by SOLPS-ITER simulation, and a schematic neutral transport model is used to interpret the simulation results. Quantities including the neutral density, neutral compression, and ionization front are analyzed, and the influences of baffles and seeding on these quantities are revealed. The TCV gas baffles separate the TCV vessel into a divertor and a main chamber region. The two regions are separated by line segments connecting the two baffle tips with the X-point, Figure 1(d). The plasma region (colored areas in Figure 1(d)) is excluded when calculating the average main chamber and divertor neutral densities. The total deuterium density is defined as $n_n \equiv 2n_{D2} + n_D$.

### 3.2.1 Observations

Most divertor neutrals are located in the reservoirs of the divertor private flux region (PFR) and common flux regions (CFR), which are separated by the plasmas of the divertor legs, Figure 6(a-d). CFR neutrals can freely transit past the SOL plasma to reach the main chamber, though the area that neutrals can transit through is considerably decreased by baffles. The averaged divertor neutral density, $\langle n_n \rangle_{div}$, significantly increases with baffles, and increases with higher nitrogen seeding rate, Figure 6(e). Without seeding, the baffled $\langle n_n \rangle_{div}$ is 2.6 times greater than the unbaffled $\langle n_n \rangle_{div}$. This difference increases further with seeding. The averaged main chamber neutral density, $\langle n_n \rangle_{main}$, decreases to a third of its unbaffled value with baffles, but does not depend on the nitrogen seeding rate, Figure 6(f).

The neutral compression $c_D$, here defined as the ratio of divertor and main chamber neutral density, $c_D = \langle n_n \rangle_{div}/\langle n_n \rangle_{main}$, is widely used in previous works studying the divertor neutral transport [27, 44, 45]. Higher $c_D$ indicates better divertor neutral confinement, which is one of the main motivations of increasing the divertor closure. Without seeding, $c_D$ increases by a factor of seven with baffles, and even more with seeding, Figure 6(g).

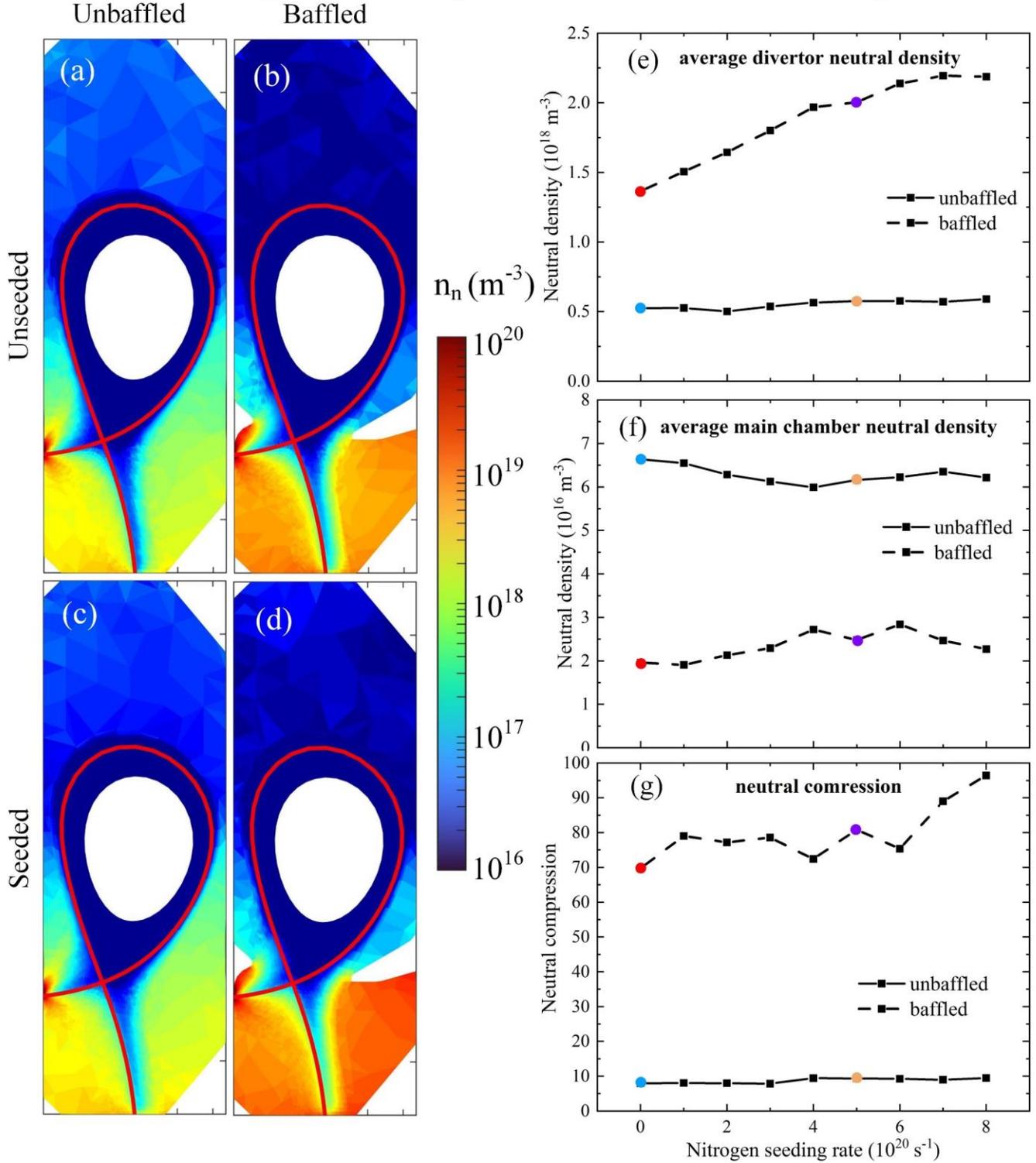

Figure 6. Changes of neutral quantities with baffles and seeding. (a)-(d) Total neutral density distribution, $n_n$, of the four cases in Table II. (a) Unbaffled, unseeded. (b) Baffled, unseeded. (c) Unbaffled, seeded. (d) Baffled, seeded. (e)-(g) Nitrogen seeding scans of the averaged neutral densities and the compression. (e) Divertor neutral density. (f) Main chamber neutral density. (g) Neutral compression. The electron density at the separatrix is fixed as $1.5 \times 10^{19}$ m$^{-3}$. The colored points correspond to the four simulation cases in Table II.

The neutral ionization distribution is affected by baffles and seeding, due to changes of the neutral density and plasma parameters. Baffles increase the ionization rate in the divertor, and extend the ionization region from the targets towards the X-point, Figure 7(a-d). Meanwhile, nitrogen seeding decreases the divertor ionization rate, and moves the ionization region away from the target towards the X-point, Figure 7(a-d).

The ionization front, defined as the poloidal position above which more than 90% of the ionization occurs, is introduced to quantify the changes of the ionization distribution. Without baffles and seeding, the ionization front is located at the target as the divertor is attached. Baffles and seeding, both, move the ionization front towards the X-point, Figure 7(e).

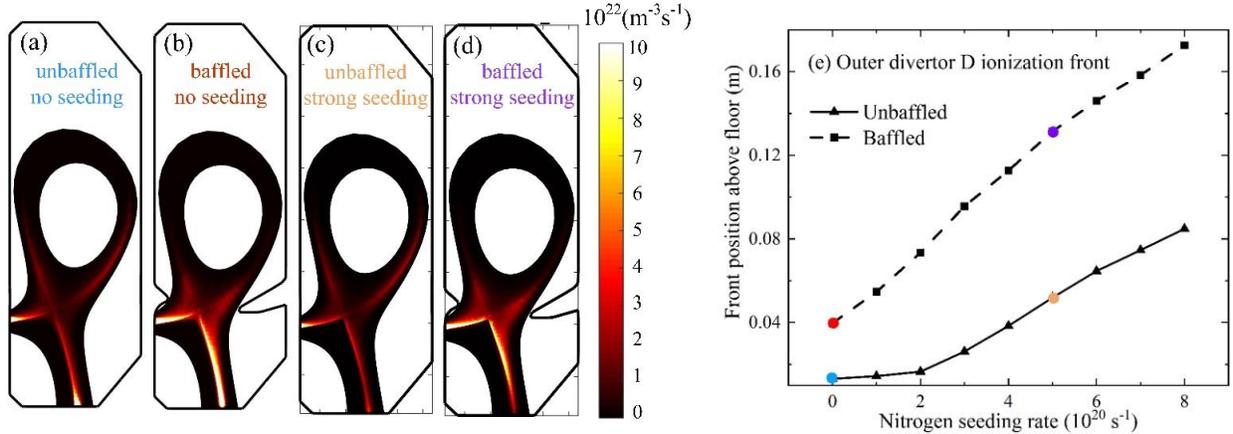

Figure 7. (a)-(d) Ionization rate distribution of the four simulation cases in Table II. The electron density at the upstream separatrix is $1.5 \times 10^{19}$ m$^{-3}$ and the nitrogen seeding rate is $5.0 \times 10^{20}$ s$^{-1}$ in the seeded cases. (e) Nitrogen seeding scan of the distance of the deuterium ionization front from the outer target. The colored points correspond to the four simulation cases in Table II.

3.2.2 Schematic model for neutral transport in a baffled divertor

The observed dependencies of neutral and ionization distributions on baffles and seeding, section 3.2.1, are explained using a simple, schematic neutral transport model [46].

The model assumes stationary condition, with perfect recycling, and no volumetric sources and sinks for neutrals. The neutrals that arise at the target from the recycling of the target ion flux, $\Phi_{i,t}$, are either directly ionized in front of the target, $S_{ion,t}$, or leak into the divertor. These divertor neutrals can then either ionize in the divertor, $S_{ion,div}$, or transit past the baffles into the main chamber region, where they ultimately also ionize, $S_{ion,main}$, as illustrated in Figure 8(a). The target ion flux, therefore, consists of three parts:

$$\Phi_{i,t} = S_{ion,t} + S_{ion,div} + S_{ion,main} \tag{8}$$

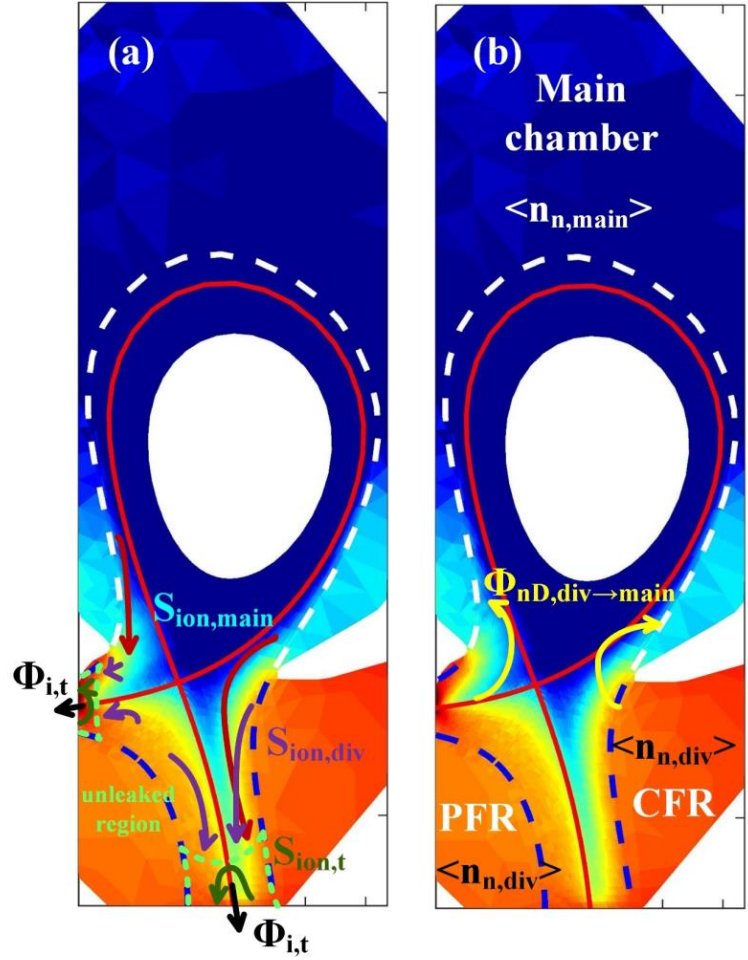

Figure 8. Schematic of the neutral transport model. (a) Three components of the target ion flux consisting of ionized divertor neutrals ($S_{ion,div}$), ionized main chamber neutrals ($S_{ion,main}$), and neutrals ionized near the target ($S_{ion,t}$) according to Equation (8). (b) Neutral density in the divertor region $\langle n_n \rangle_{div}$, main chamber region $\langle n_n \rangle_{main}$, and neutral flux towards the main chamber region $\Phi_{n,div \to main}$.

Note that the last two terms on the RHS of Equation (8) represent neutrals that do not ionize directly in front of the target (i.e. the region marked in green in Figure 8(a)) and leak into the divertor, and potentially, the main chamber. They can be characterized by a leakage factor $f_{leak}$, which depends on the target plasma parameters and the local geometry. Equation (8) is rewritten with $f_{leak}$ as follows:

$$\Phi_{i,t} = S_{ion,t} + f_{leak}\Phi_{i,t} \tag{9}$$

In this model, the neutral particle sinks due to ionization in the divertor and main chamber are assumed to be proportional to the respective neutral densities. The neutral sink due to ionization by plasma can be characterized by an effective neutral pumping speed, with the divertor and main chamber plasma pumping speeds $S_{div}$ and $S_{main}$:

$$S_{ion,div} = \langle n_n \rangle_{div} S_{div} \tag{10}$$

$$S_{ion,main} = \langle n_n \rangle_{main} S_{main} \tag{11}$$

The neutral flux from the divertor to the main chamber, sketched in Figure 8(b), is proportional to the neutral density difference between the two regions and can therefore be characterized by a neutral conductance from divertor to main chamber, $C_{cond}$:

$$\Phi_{n,div \to main} = C_{cond}(\langle n_n \rangle_{div} - \langle n_n \rangle_{main}) \tag{12}$$

Assuming $\langle n_n \rangle_{div} \gg \langle n_n \rangle_{main}$ and that the neutral flux, $\Phi_{n,div \to main}$, is returned as an ion flux, $S_{ion,main}$, Equations (11) and (12) yield an expression for the neutral compression:

$$c_D = \frac{S_{main}}{C_{cond}} \tag{13}$$

Equation (13) indicates that $c_D$ is inversely proportional to the neutral conductance from the divertor to the main chamber region. The model, thereby, explains the increase of $c_D$ with baffling. Combining Equations (8)-(12), and again assuming $\langle n_n \rangle_{div} \gg \langle n_n \rangle_{main}$ in Equation (12), the corresponding divertor neutral density is:

$$\langle n_n \rangle_{div} = \frac{f_{leak} \Phi_{i,tar}}{S_{div} + C_{cond}} \tag{14}$$

Equation (14) shows that the divertor neutral density increases with baffles, i.e. decreasing neutral conductance from divertor into the main chamber, $C_{cond}$. This increase will, however, saturate once $C_{cond}$ is small compared to $S_{div}$.

The factors $f_{leak}$, $S_{div}$, and $C_{cond}$ and their dependence on baffles and seeding can be calculated from the SOLPS-ITER simulations, as discussed in the following. The calculations of these quantities are based on first-principle physical quantities from the simulations, such as the ionization distribution, neutral flux and neutral density.

The procedure of the outer divertor leakage factor calculation is briefly presented below. For a given position at the outer targe, $p_t$, the scan of recycled neutral trajectories with emission angle θ ranging between 0 and π is performed. For each neutral trajectory, $s$, the ratio of the differential neutral trajectory, d$s$, and the deuterium neutral ionization mean free path, $\lambda_{Di}$, is integrated between the emission location and the trajectory's intersection with the outmost plasma grid, from which $f_{leak}$ of this neutral trajectory is calculated as follows.

$$f_{leak}(p_t, \theta) = \exp\left(-\int_{s_{start}}^{s_{end}} \frac{ds}{\lambda_{Di}(p_t,\theta)}\right) \tag{15}$$

Note that for trajectories that cross the core boundary, $f_{leak}$ is assumed to be zero, i.e. all neutrals ionize. Charge exchange collisions are currently neglected. To calculate the total leakage factor, $f_{leak}(p,\theta)$ is first averaged over the neutral emission angle range, where a cosine distribution is assumed, and then averaged over the extent of the target, while weighted with the target deuterium ion flux:

$$f_{leak} = \frac{1}{\Phi_{i,t}} \int_{t_{start}}^{t_{end}} \int_0^\pi \Gamma_{\perp,t}(p) f_{leak}(p,\theta) \frac{\cos(0.5\pi - \theta)}{2} d\theta \, dA(p) \tag{16}$$

Here $\Gamma_{\perp,t}$ is the perpendicular target ion flux density, $\Phi_{i,t}$ is the total target ion flux, $dA$ is the differential target area.

The divertor plasma pumping speed, $S_{div}$, is calculated from Equation (10), using the divertor ionization rate and the average neutral density from SOLPS-ITER. The neutral conductance from the divertor to the main chamber, $C_{cond}$, is calculated from Equation (12), using the neutral flux crossing the interface between the divertor and the main chamber, $\langle n_n \rangle_{div}$ and $\langle n_n \rangle_{main}$.

The simulations show that baffles decrease $C_{cond}$ down to a third of its unbaffled value, Figure 9(a). This large decrease is primarily caused by the reduction of the area that neutrals have to transit from the divertor past the baffles to the main chamber. They also reveal a small decrease of $C_{cond}$ with increasing seeding rate, presumably due to cooler and, hence slower, divertor neutrals, an effect that likely contributes to the reduction of $C_{cond}$ with baffles, too. The divertor plasma pumping speed, $S_{div}$, also decreases with baffles by approximately 50% at all seeding levels, Figure 9(b). $S_{div}$ also decreases with increasing seeding rate. The decrease of $S_{div}$ with baffles and seeding could be caused by lower divertor plasma temperatures increasing the mean free path for ionization as well as by cooler divertor neutrals with baffles and seeding. The calculated leakage factor increases with both baffles and seeding, Figure 9(c). Without seeding, $f_{leak}$ with baffles is twice as high than without baffles. $f_{leak}$ increases faster without baffles than with baffles, leading to less of a difference at high seedings rates, Figure 9(c).

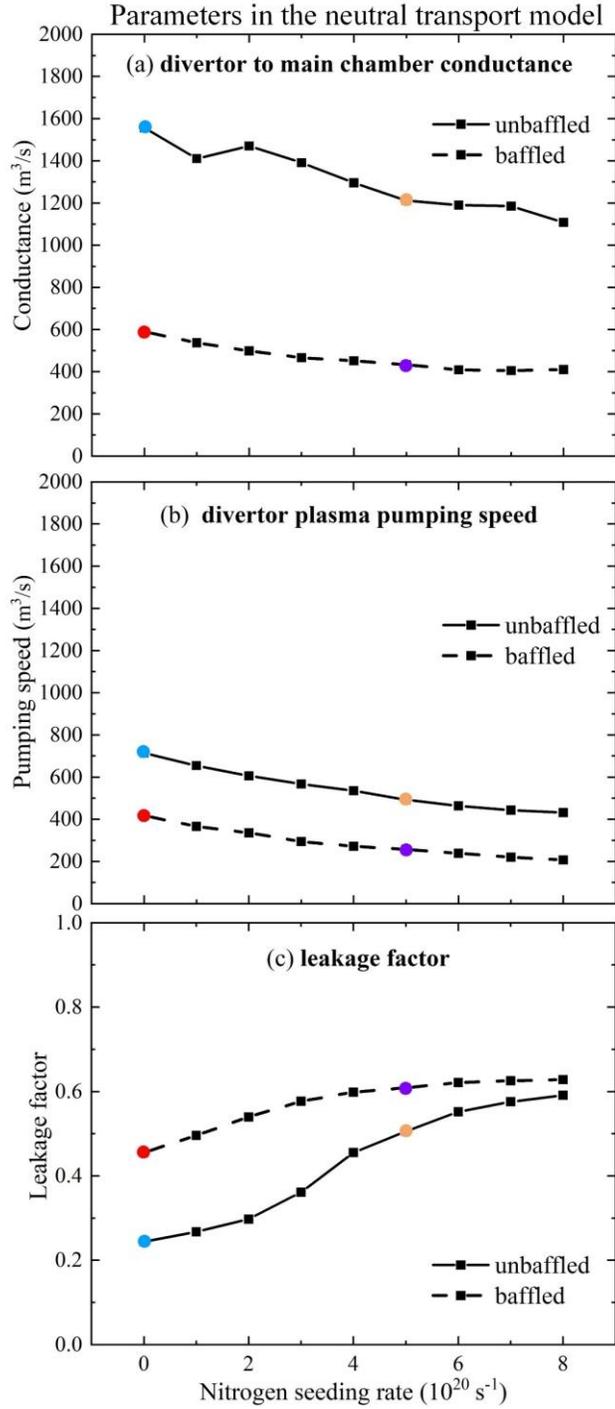

Figure 9. Dependence of the terms in the neutral transport model on baffling and seeding. (a) Divertor to main chamber neutral conductance $C_{cond}$. (b) Divertor plasma pumping speed $S_{div}$. (c) Leakage factor $f_{leak}$. The electron density at the separatrix is fixed as $1.5 \times 10^{19}$ m$^{-3}$. The colored points correspond to the four simulation cases in Table II.

The effect of baffles and seeding on $\langle n_n \rangle_{div}$, $\langle n_n \rangle_{main}$, $c_D$, and the ionization front observed in section 3.2.1 are explained by the proposed neutral transport model. The increase of $\langle n_n \rangle_{div}$ with baffles is primarily due to the lower $C_{cond}$. Consequently, the divertor plasma becomes denser and colder, due to higher plasma-neutral collisionality. This leads to higher $f_{leak}$ as the ionization front moves away from the target, higher target ion flux (Figure 2(d)), and lower $S_{div}$ due to colder divertor neutrals. Baffles decrease $\langle n_n \rangle_{main}$ because the increase of $\langle n_n \rangle_{div}$ is less than the increase of $\Phi_{n,div \to main}/C_{cond}$ (not shown). Higher $\langle n_n \rangle_{div}$ and lower $\langle n_n \rangle_{main}$ with baffles together increase $c_D$. Contrary to the baffles, nitrogen seeding does not considerably affect the neutral conductance, but cools down the divertor plasmas with stronger impurity radiation, which increases $f_{leak}$, and, therefore, increases $\langle n_n \rangle_{div}$.

For the simulated configurations $S_{div}$ is of the same order of magnitude as $C_{cond}$ for baffled and unbaffled divertor, with the ratio $C_{cond}/S_{div}$ being lower with baffles, Figure 9(a-b). With tighter baffling, it is possible that $C_{cond} \ll S_{div}$, such that further increase of the divertor closure can no longer increase $\langle n_n \rangle_{div}$. The trend that $c_D$ increases with lower $C_{cond}$ is unaffected by tighter baffling, according to Equation (13), where further increase of $\dot{c}_D$ should primarily be driven by lower $\langle n_n \rangle_{main}$, rather than higher $\langle n_n \rangle_{div}$.

### 3.3. Impurity retention and radiation

Baffles decrease the impurity neutral conductance from the divertor to the main chamber, similar to their effect on deuterium neutrals. Baffles also change the divertor D$^+$ flow pattern, and therefore, the friction force between D$^+$ and impurities. In addition, baffles increase the target D$^+$ flux and, hence, the source of sputtered carbon. Consequently, the distributions of impurity neutral density, ion density, and radiated power, change with baffles. These effects are analyzed to validate the compatibility of baffles with nitrogen seeding, and their potential synergies on achieving divertor detachment.

Since both impurity neutrals and ions radiate, the *impurity retention* is here defined as the ratio of average total impurity density in the divertor and in the main chamber region:

$$R_{imp} = \frac{\langle n_{imp} \rangle_{div}}{\langle n_{imp} \rangle_{main}} \tag{17}$$

with $n_{imp}$ being the total impurity density (including neutral impurities and all ionized states), and the index *imp* specifying the impurity species (*C* or *N* for carbon or nitrogen, respectively). The separation of the main chamber and divertor region is the same for the neutral compression. The impurity retention in the divertor varies with baffling, the impurity species, and the nitrogen seeding rate.

3.3.1. Analysis of nitrogen retention

Since the nitrogen recycling coefficient is well below unity, wall-pumped nitrogen is replaced with a continuous nitrogen flow from a seeding valve, which is here located in the PFR of the divertor, Figure 1(b). Most seeded neutral nitrogen must, therefore, first transit through the divertor legs into the CFR, before it can be transported past the baffles into the main chamber. This leads to a considerable neutral nitrogen density difference between the PFR, the divertor CFR, and the main chamber, i.e. $\langle n_{n,N} \rangle_{div,PFR} \gg \langle n_{n,N} \rangle_{div,CFR} \gg \langle n_{n,N} \rangle_{main}$, Figure 10 and Figure 11(a-c). Baffles considerably increase $\langle n_{n,N} \rangle_{div,CFR}$, Figure 11(b), via two mechanisms. First, the divertor to main chamber neutral nitrogen conductance, $C_{cond,N}$, decreases with baffles, similar to neutral deuterium. Second, baffles cool down the divertor plasma, increasing the transparency of the divertor legs for neutral nitrogen and facilitating the neutral nitrogen transport from the divertor PFR to the CFR. The divertor plasma transparency also increases with seeding, causing $\langle n_{n,N} \rangle_{div,CFR}$ to increase faster than linearly with the seeding. The increased neutral nitrogen flux from the PFR into the CFR is, however, too small to affect $\langle n_{n,N} \rangle_{div,PFR}$. As a result, $\langle n_{n,N} \rangle_{div,PFR}$ is predominantly determined by the balance of the seeding gas source in the divertor PFR and the ionization of PFR neutrals, and, therefore, unaffected by baffles, increasing linearly with seeding. Despite the higher value of $\langle n_{n,N} \rangle_{div,CFR}$, baffles decrease $\langle n_{n,N} \rangle_{main}$ due to the lower $C_{cond,N}$, Figure 11(a, c).

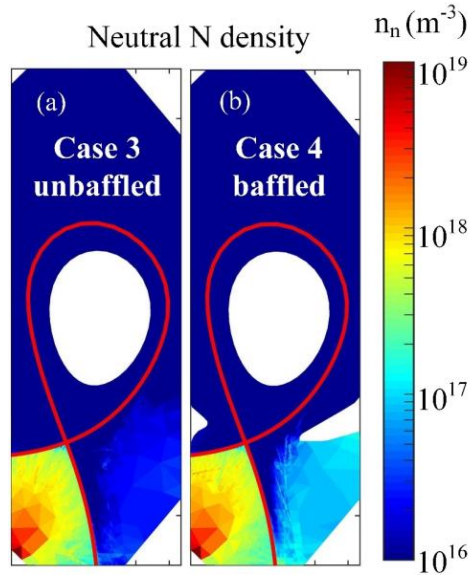

Figure 10. Neutral nitrogen distribution without and with baffles and with seeding (case 3 and 4 of Table II). (a) Neutral nitrogen density without baffles. (b) Neutral nitrogen density with baffles. The electron density at the upstream separatrix is $1.5 \times 10^{19}$ m$^{-3}$ and the nitrogen seeding rate is $5 \times 10^{20}$ s$^{-1}$ in the seeded cases.

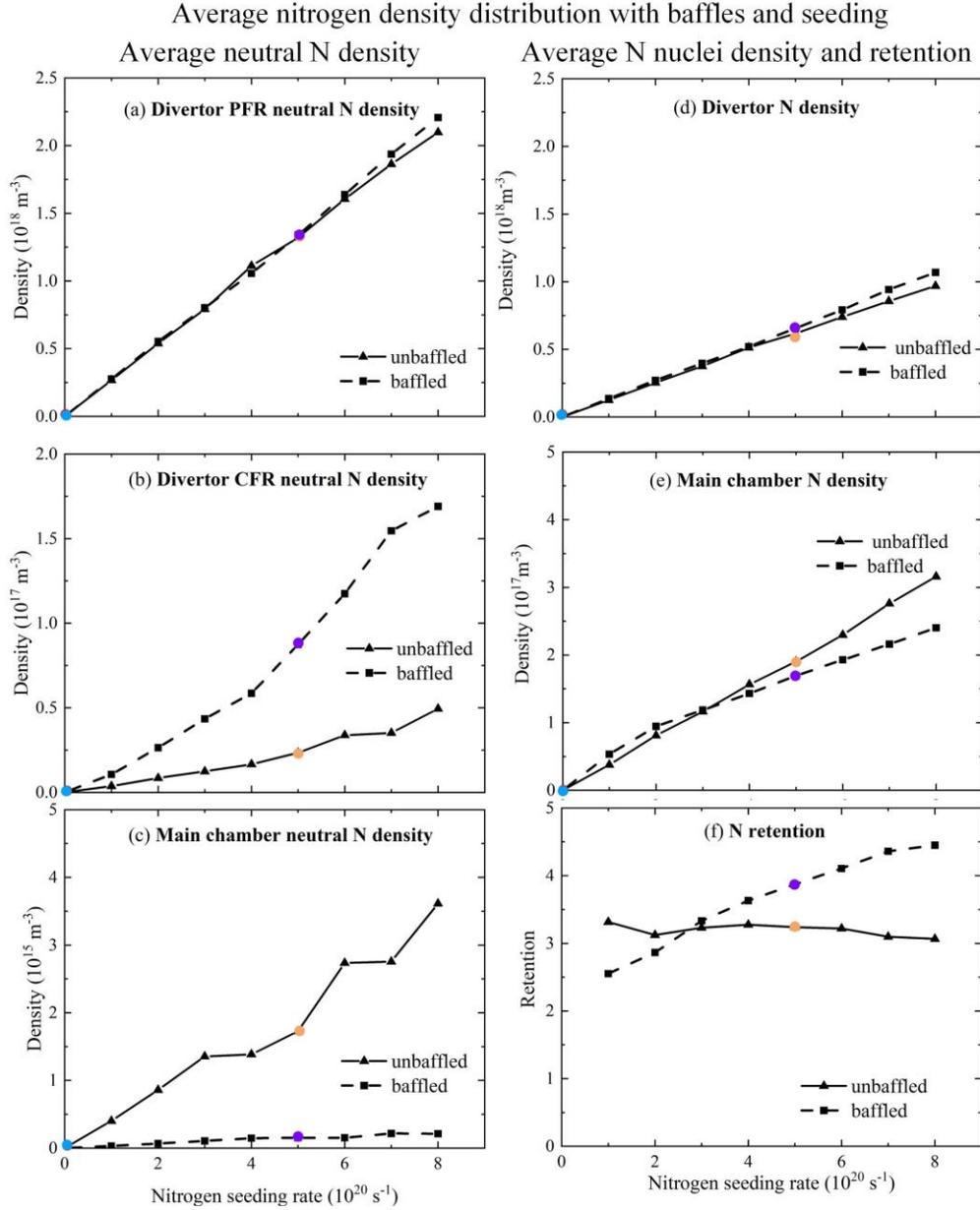

Figure 11. Average neutral nitrogen density in (a) Divertor PFR. (b) Divertor CFR. (c) Main chamber. Average nitrogen nuclei density in (d) Divertor. (e) Main chamber. The nitrogen retention (f). The electron density at upstream separatrix is fixed as $1.5 \times 10^{19}$ m$^{-3}$ and the colored points correspond to the four simulation cases in Table II.

The nitrogen nuclei density, and the nitrogen retention, are affected by both, nitrogen neutral and nitrogen ion densities. Divertor neutral nitrogen, mostly from the divertor PFR neutral, contributes more than 50% to $\langle n_N \rangle_{div}$. Like $\langle n_{n,N} \rangle_{div,PFR}$, $\langle n_N \rangle_{div}$ increase linearly with the seeding rate, and is barely affected by baffles, Figure 11(a, d). Meanwhile, neutral nitrogen contributes less than 1% to $\langle n_{n,N} \rangle_{main}$ at all seeding levels in both, unbaffled and baffled divertors. Therefore, the observed trends of $\langle n_N \rangle_{main}$ are due to a change of nitrogen ion leakage flux from divertor to the main chamber. At low seeding rates $\langle n_N \rangle_{main}$ increases faster with the seeding rate with baffles, Figure 11(e). This increase of nitrogen ion leakage flux in the presence of the baffles is due to a change of the D$^+$ flow pattern, which moves the ion stagnation point closer to the target (not shown) and, thereby, promotes nitrogen ion leakage into the main chamber. As the seeding rate increases, the dependence of $\langle n_N \rangle_{main}$ on seeding weakens with baffles. Since these divertor plasmas feature the lowest temperatures, the decreased ion thermal force decreases the ion leakage and slows the increase of $\langle n_N \rangle_{main}$ with the seeding rate.

These dependencies of $\langle n_N \rangle_{div}$ and $\langle n_N \rangle_{main}$ lead to a slightly lower $R_N$ with baffle at low seeding levels, but a ~50% higher $R_N$ with baffles than without baffles at highest simulated seeding levels, Figure 11(f).

3.3.2. Analysis of the carbon retention

Carbon is mainly sourced by sputtering at the target, with C leaking into, both, PFR and CFR. The neutral carbon density, therefore, exhibits smaller difference between the divertor PFR and CFR than neutral nitrogen, similar to neutral deuterium, i.e. $\langle n_{n,C} \rangle_{div,PFR} \sim \langle n_{n,C} \rangle_{div,CFR}$, Figure 12 and Figure 13(a-c). Baffles decrease the divertor to main chamber carbon neutral conductance, $C_{cond,C}$, and increase the target particle flux, both leading to higher $\langle n_{n,C} \rangle_{div,PFR}$ and $\langle n_{n,C} \rangle_{div,CFR}$. Seeding decreases the target particle flux, therefore slightly decreases $\langle n_{n,C} \rangle_{div,PFR}$ and $\langle n_{n,C} \rangle_{div,CFR}$, Figure 13(a, b). Baffles decrease $\langle n_{n,C} \rangle_{main}$ due to lower $C_{cond,C}$, Figure 13(c).

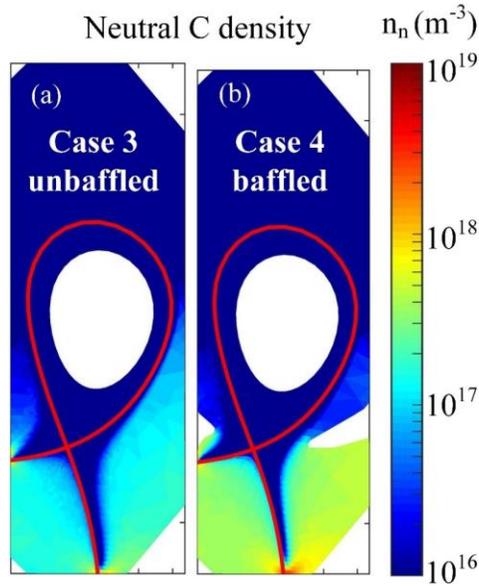

Figure 12. Neutral carbon distribution without and with baffles and with seeding (case 3 and 4 of Table II). (a) Neutral carbon density without baffles. (b) Neutral carbon density with baffles. The electron density at the upstream separatrix is $1.5 \times 10^{19}$ m$^{-3}$ and the nitrogen seeding rate is $5 \times 10^{20}$ s$^{-1}$.

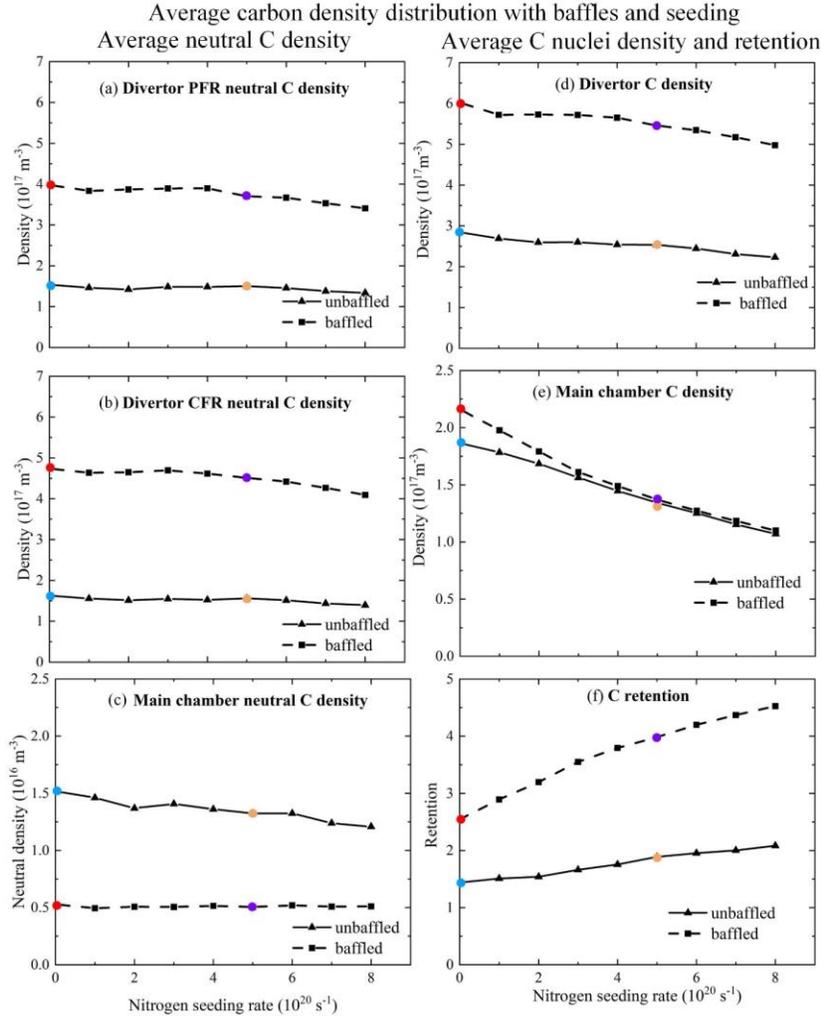

Figure 13. Average neutral carbon density in (a) Divertor PFR. (b) Divertor CFR. (c) Main chamber. Average carbon nuclei density in (d) Divertor. (e) Main chamber. The carbon retention (f). The electron density at the upstream separatrix is fixed as $1.5 \times 10^{19}$ m$^{-3}$ and the colored points correspond to the four simulation cases in Table II.

Similar to nitrogen, divertor neutral carbon contributes more than 50% to the divertor carbon nuclei density $\langle n_C \rangle_{div}$. $\langle n_C \rangle_{div}$ is, therefore, also higher with baffles, and slightly decreases with seeding, Figure 13(d). Also similar to nitrogen, neutral carbon contributes little, less than 10%, to $\langle n_C \rangle_{main}$ and observed trends of $\langle n_C \rangle_{main}$ are due to changes of the carbon ion leakage flux from the divertor to the main chamber. Baffles cause higher target particle flux, more carbon source, and more leaked carbon ions into the main chamber, which increase $\langle n_C \rangle_{main}$. $\langle n_C \rangle_{main}$ is lower with seeding due to lower target particle flux and less carbon source, Figure 13(e).

These dependencies of $\langle n_C \rangle_{div}$ and $\langle n_C \rangle_{main}$ lead to higher $R_C$ with baffles, which generally increases with higher nitrogen seeding rates, Figure 13(f).

3.3.3. Analyses of the impurity radiation

Since the impurity-radiated power is proportional to the impurity density, the trends of nitrogen and carbon density distributions lead to pertinent changes in the radiated power distributions. Overall, the effects of baffles and seeding on nitrogen and carbon radiation distributions are consistent with their effects on the impurity density distributions, Figure 14.

The nitrogen radiation increases with higher seeding rate. Baffles do not change the divertor nitrogen radiation, while the core nitrogen radiation is significantly lower at high seeding level with baffles, Figure 14(e, g), (f, h). The carbon radiation decreases with higher seeding rate. Baffles increase the carbon radiation especially in the divertor region, Figure 14(a, c), (b, d).

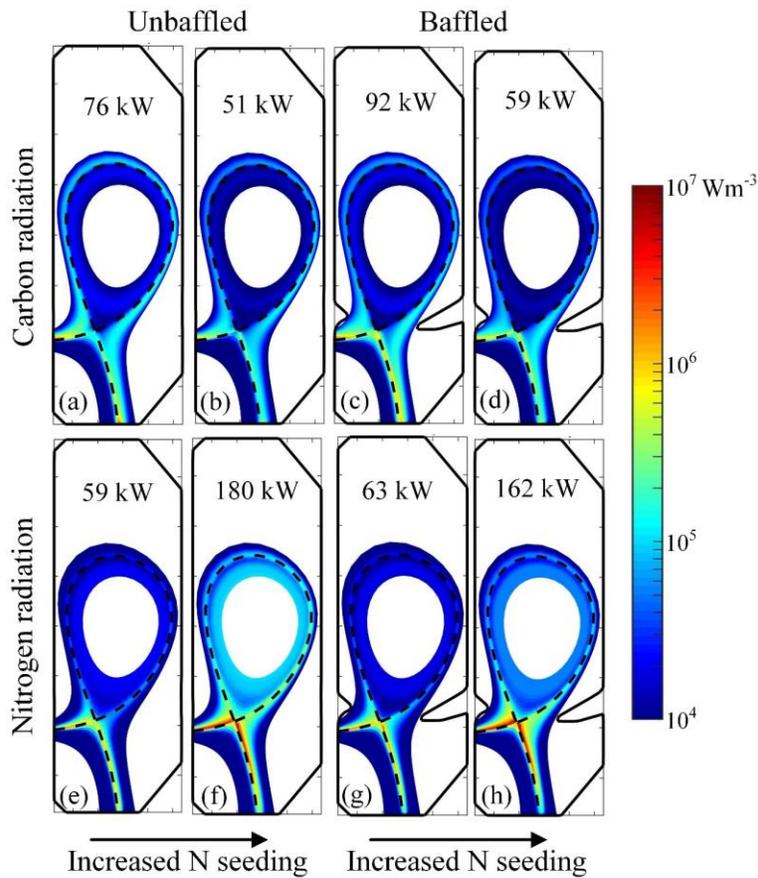

Figure 14. Impurity radiation distribution. (a)-(d) Carbon radiation. (e)-(h) Nitrogen radiation. The nitrogen seeding rate is increased from 2 × $10^{20}$ atom/s to 8 × $10^{20}$ atom/s. The electron density at the separatrix is fixed as 1.5 × $10^{19}$ $m^{-3}$. The total radiated power of carbon or nitrogen in the simulated region is shown for each case.

The radiated power is integrated in the divertor and main chamber region. Adding baffles has only a negligible effect on the divertor nitrogen radiation, but considerably decreases the main chamber nitrogen radiation, Figure 15(a, b), which is consistent with the effects of baffles on $\langle n_N \rangle_{main}$ and $\langle n_N \rangle_{div}$. At the highest seeding rate divertor radiation contributes 54% to the total nitrogen radiation, which increases to 64% with baffles, Figure 15(a, b).

Baffles greatly increase the divertor carbon radiation, but have less impact on the main chamber carbon radiation, Figure 15(c, d), which is consistent with the effects of baffles on $\langle n_C \rangle_{main}$ and $\langle n_C \rangle_{div}$. At highest seeding rate divertor carbon radiation contributes 56% of the total carbon radiation, which increases to 71% with baffles, Figure 15(c, d). Baffles also increase the total carbon radiation by up to 28% due to increased carbon sources at the targets caused by higher target particle flux. Nitrogen seeding decreases the overall carbon radiation due to reduced carbon sources at the targets.

Radiation from all simulated species, including carbon, nitrogen and deuterium, are summed up and analyzed, Figure 15(e, f). The contribution of deuterium radiation to the total radiation decreases with seeding and increases with baffles, but generally remains small ranging from 3% to 17%. The total radiation of all species increases with baffles at low seeding levels, with the effect less important at high seeding levels. Divertor radiation of all species is always slightly higher with baffles, and the main chamber radiation of all species is always considerably lower with baffles. The divertor and main chamber radiation of all species trivially increases with seeding.

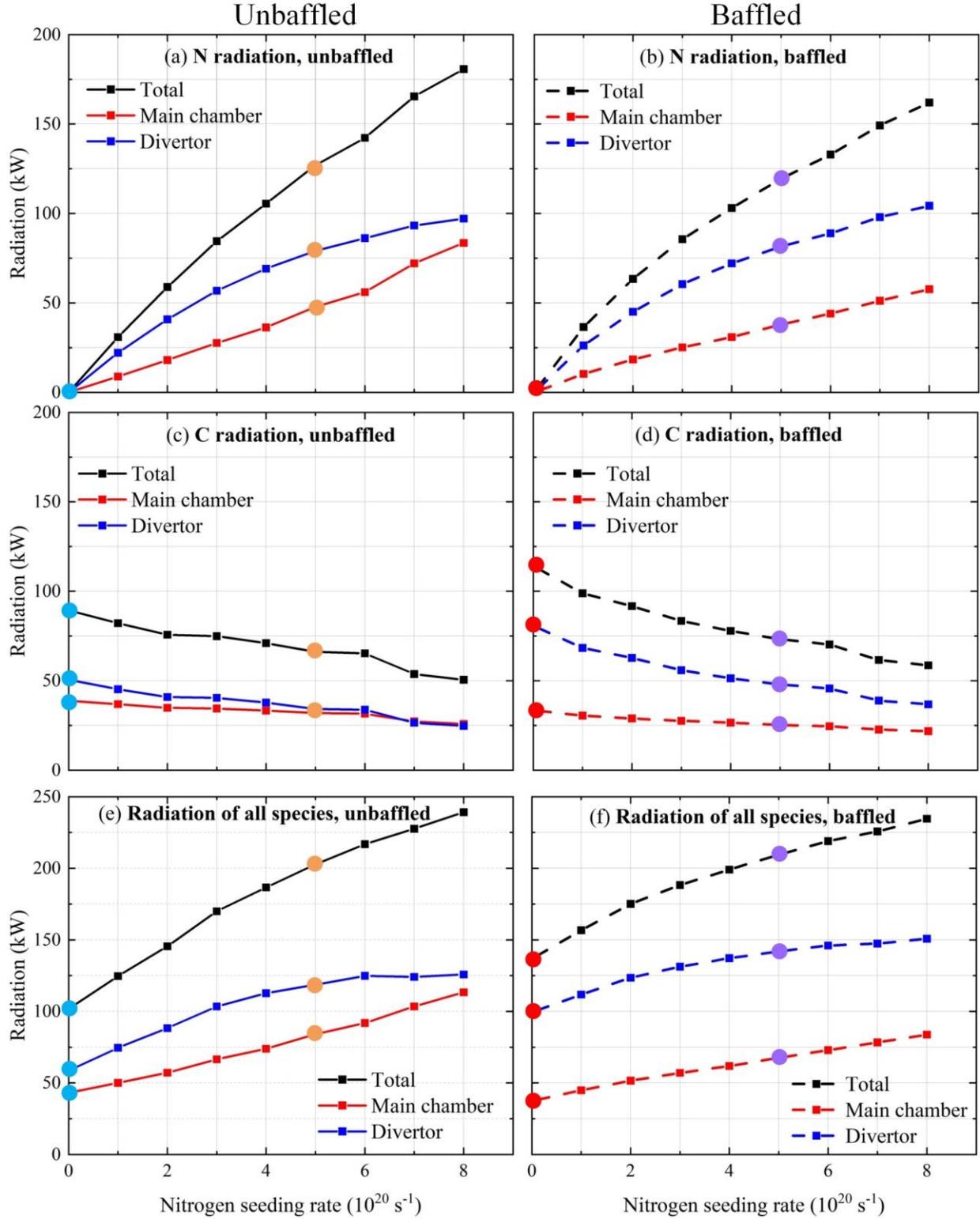

Figure 15. Nitrogen seeding scan of impurity radiation in the simulated divertor and the main chamber regions. (a)-(b) Nitrogen radiation. (c)-(d) Carbon radiation. (e)-(f) Total radiation of carbon, nitrogen and deuterium. Region definition is shown in Figure 1(d). The electron density at the upstream separatrix is fixed as $1.5 \times 10^{19}$ m$^{-3}$ and the colored points correspond to the four simulation cases in Table II.

### 3.3.4. Analyses of the seeding operational window

While impurity seeding increases the divertor power dissipation, the seeding rate is limited by the core performance, which degrades with excessive core impurity concentration. The operational window that features a sufficiently large reduction of poloidal heat fluxes due to impurity radiation, without an excessive core impurity concentration, and its scaling towards a reactor, is a critical open issue. Above analyses illustrate that baffles can decrease the core nitrogen density and radiation, suggesting a potential synergy between baffles and seeding. The operational window for nitrogen seeding with and without baffles is discussed below to demonstrate such synergy.

The upstream impurity concentration is defined as the ratio of total impurity ion density (here carbon and nitrogen) to electron density, at the outboard midplane separatrix, $c_{imp}^{sep} = \langle \frac{\sum_i n_{C,N}^{i+}}{n_e} \rangle_{sep}$. The operational window of seeded detachment is defined as the conditions where the peak outer target temperature is lower than 5 eV, and $c_{imp}^{sep}$ is within tolerable levels, Figure 16. While a maximum tolerable $c_{imp}^{sep}$ is difficult to define, lower values are generally more favorable. The baffled divertor features significantly lower core pollution for the same plasma temperature at the targets than the unbaffled divertor. Higher input power shifts the $c_{imp}^{sep} - T_{ot,max}$ to higher $T_{ot,max}$ (not shown). The analyses of the seeding operational window show that baffles can significantly decrease the upstream nitrogen concentration while keeping the divertor detached, which widens the seeding operational window.

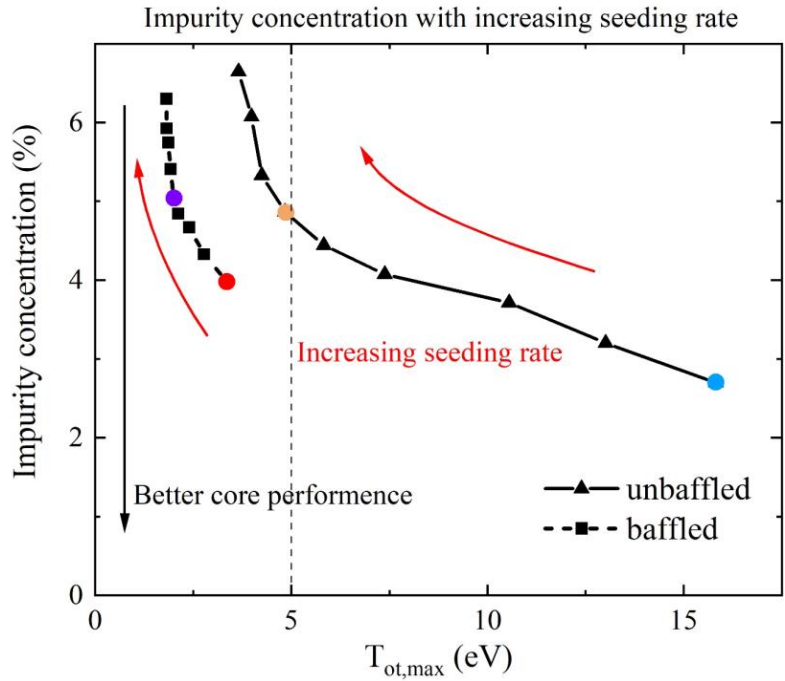

Figure 16. The impurity concentration at the separatrix as function of peak outer target temperature, obtained with nitrogen seeding rate from 0 to $8\times10^{20}$ s$^{-1}$. The electron density at separatrix is fixed as $1.5 \times 10^{19}$ m$^{-3}$ and the colored points correspond to the four simulation cases in Table II. Detachment threshold is marked by the vertical dashed line.

4. COMPARISON WITH TCV EXPERIMENTS

In this section, SOLPS-ITER simulations are compared with TCV experiments. The comparison seeks to improve the understanding of experimental trends and identify potential reasons for discrepancies, rather than aiming for quantitative agreement. Key predictions to verify in TCV are:

1) Target parameters: Baffles increase the target electron density and particle flux, and decrease the target electron temperature and heat flux. Seeding decreases the target electron temperature, particle flux and heat flux, but does not change the target electron density.

2) Neutral density and compression: Baffles increase the divertor neutral density and neutral compression, and decrease the main chamber neutral density.

3) Impurity radiation: Seeding increases the total radiation of all plasma species increases, while baffling only increases the total radiation for low seeding rates. Radiation of all plasma species in the divertor increases with both baffles and seeding. Divertor carbon radiation increases with baffles and decrease with seeding. Divertor nitrogen radiation is not affected by baffles, and increase with seeding.

## 4.1. Criteria for comparison

The comparison is based on Ohmic L-mode TCV discharges with a plasma current of 250 kA (consistent with the simulations in Section 3) and line-averaged electron densities of approximately $3.7\pm0.2\times10^{19}$ m$^{-3}$. The SOLPS upstream electron density, $n_{e,sep}=1.35\times10^{19}$m$^{-3}$, is matched with the Thomson scattering measurement of these discharges. Nitrogen seeding starts at 0.85s, with a linearly increasing rate until a disruption occurs, Figure 17. While the radiated power and seeding rate at the disruption vary among identically programmed discharges, all discharges show an increase of the radiation with increasing N$_2$ seeding rate.

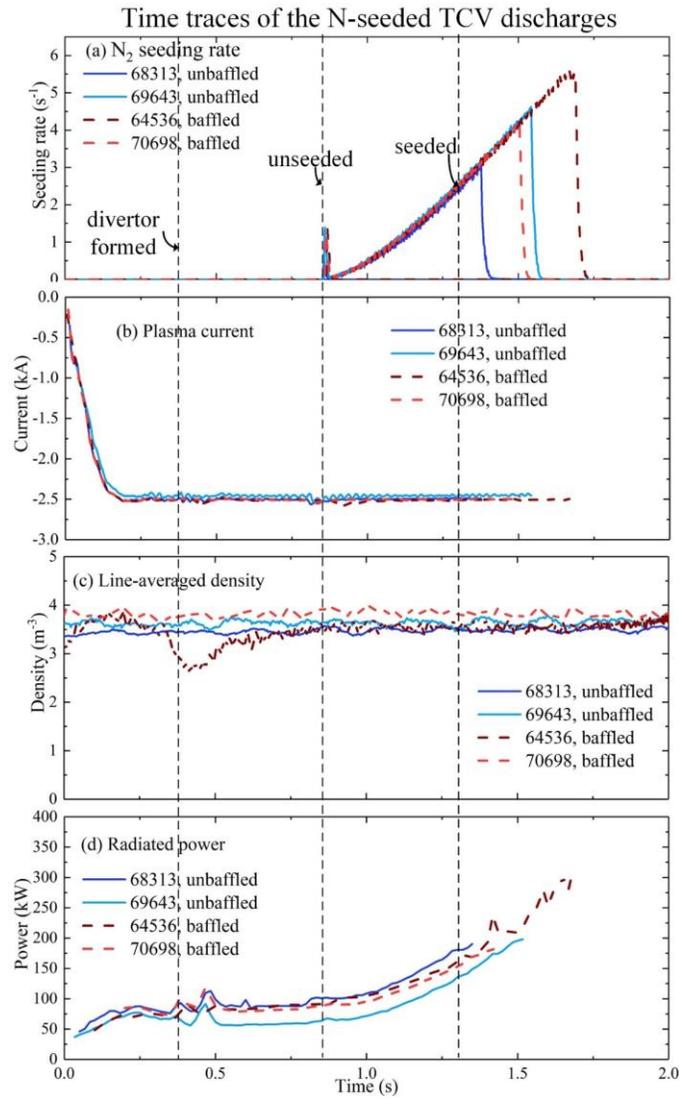

Figure 17. (a) Nitrogen seeding rate, (b) plasma current, (c) line-averaged density and (d) radiated power in unbaffled (solid) and baffled (dashed) TCV discharges. The divertor formation is marked by a first dashed line. The times that correspond to the unseeded and seeded

($5\times10^{20}$ atom/s) cases in the comparison are marked by another two dashed lines. The low radiated power of discharge 69643 is partially due to a missing upper lateral bolometer camera.

The nitrogen seeding rate is used as ordering parameter for the comparison of simulations and experiments. This approach assumes that the seeding rate ramp is sufficiently slow to result in quasi-stationary plasma states that can be compared to stationary SOLPS simulations. While faster seeding ramps lead to a systematic error, they should still show the same trends observed in simulation.

4.2. Comparison of target profiles

The outer target electron density, $n_{e,ot}$, temperature, $T_{e,ot}$, and particle flux, $\Gamma_{ot}$, are measured by wall-embedded Langmuir probes (LP) [47]. The heat flux, $q_{ot}$, is measured by an infrared thermography diagnostic (IR) [48]. The wall-embedded Langmuir probes directly measure the target ion saturation currents $J_{ot} = e\Gamma_{ot}$, and their I-V curves provide estimates of the $T_{e,ot}$.

The target ion saturation current, $J_{ot} = e\Gamma_{ot}$, is a measurement of the particle flux and reflects the difference between volumetric ion sources and sinks. The observed increase of $J_{ot}$ with baffles and decrease with seeding is consistent with the SOLPS predictions, Figure 2 and Figure 18. In all cases, SOLPS overestimates the peak of $J_{ot}$ by approximately a factor of two and predicts broader $J_{ot}$ profiles. These overestimations indicate an overestimated ionization source, which is also reflected in an overestimated divertor neutral pressure, to be discussed in Section 4.3. The double-peak in the measured $J_{ot}$ profile is not seen in these simulations, likely due to the omission of particle drifts. The double-peaked target density profile with drifts activated were previously observed in TCV simulations with UEDGE [49], and SOLPS-ITER [50].

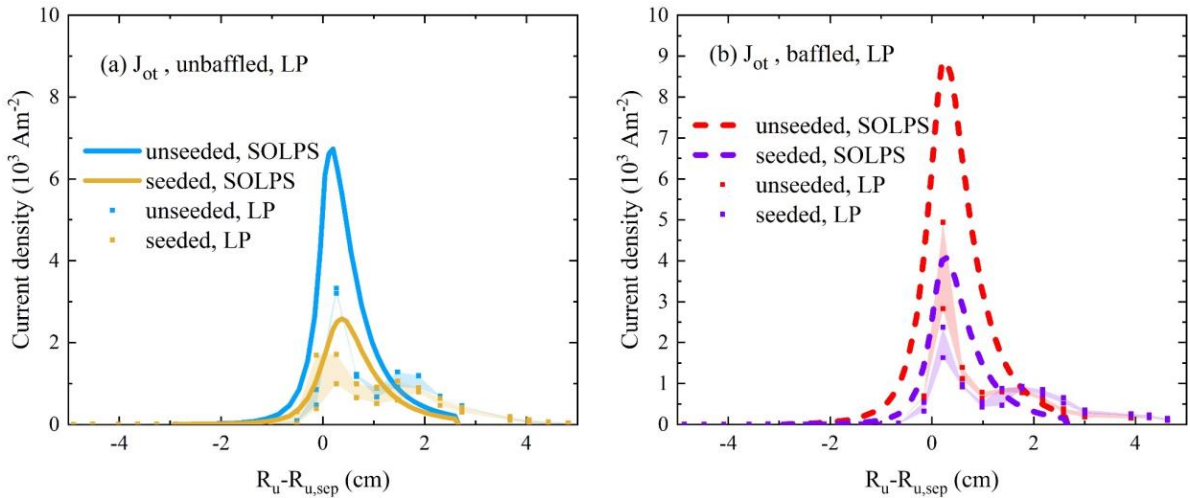

Figure 18. Comparison of measured and predicted outer particle currents in the (a) unbaffled and (b) baffled TCV configurations. The currents are taken from wall-embedded Langmuir probes, and seeding rate is $5.0 \times 10^{20}$ s$^{-1}$ in the seeded cases. Shaded areas represent the upper and lower limit of the LP measurements for the four TCV discharges shown in section 4.1. Each point of the measurement corresponds to the current and $R_u - R_{u,sep}$ of one particular probe averaged over a time interval of 5 ms.

Target electron density and temperature are key parameters which strongly influence the reaction rates in the divertor. Matching the simulation target density and temperature with the TCV experiment is, therefore, essential for reliable predictions. The decrease of target temperature by both seeding and baffles is shown in both SOLPS predictions and LP measurement, Figure 19. The lowest target temperature with seeding and baffles is well reproduced in LP measurement. The absolute values of the SOLPS predictions for the unbaffled divertor are consistent with the LP measurement, whereas the effects of baffles on target temperature is overestimated.

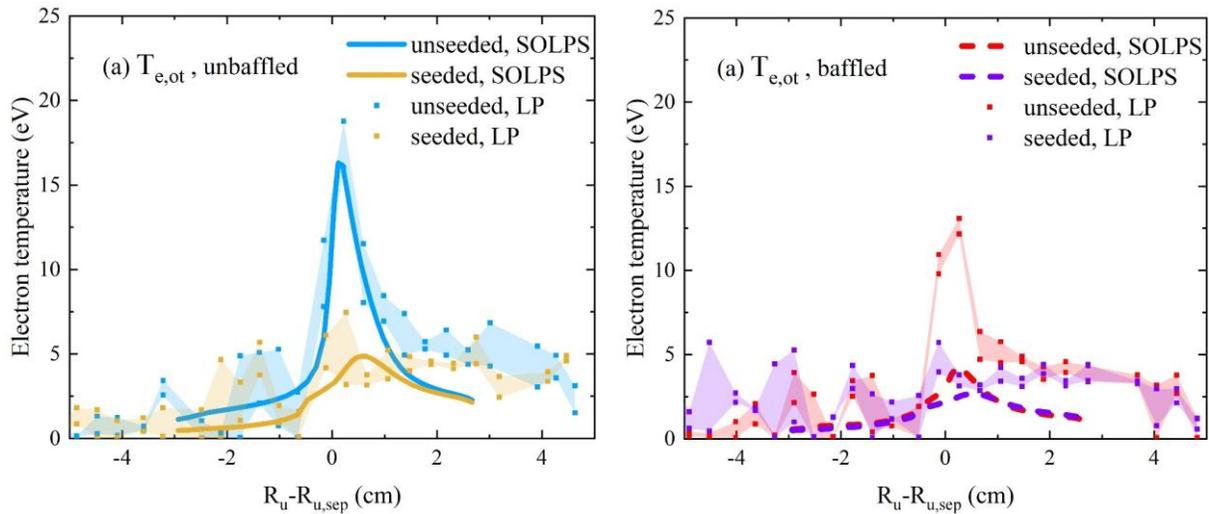

Figure 19. Comparison of measured and predicted outer target electron temperatures in the (a) the unbaffled and (b) baffled TCV configurations. Temperature profiles are obtained from wall-embedded Langmuir probes, and seeding rate is $5.0 \times 10^{20}$ s$^{-1}$ in the seeded cases.

Shaded areas represent the upper and lower limit of the LP measurements for the four TCV discharges shown in section 4.1. Each point of the measurement corresponds to the temperature and $R_u - R_{u,sep}$ of one particular probe averaged over a time interval of 5 ms.

The target electron density is derived from Langmuir probes, based on $n_t = \frac{J_t}{ec_{s,t}}$, with $J_t$ the saturation current density and $c_{s,t}$ the ion sound speed. The increase of target electron density by baffles is observed in both simulation and experiment, with SOLPS predictions higher than the LP measurements for both baffled and unbaffled conditions, Figure 20, as already observed in previous work that included drifts [38]. The trend that the target electron density is insensitive to the seeding is also observed in both simulation and experiment, Figure 20. However, the "double peak" in the measured density profile is not recovered in the simulations and may require the inclusion of drifts. The simulations also generally overestimate $n_{ot}$, consistent with the general overestimation of $\Gamma_{ot}$.

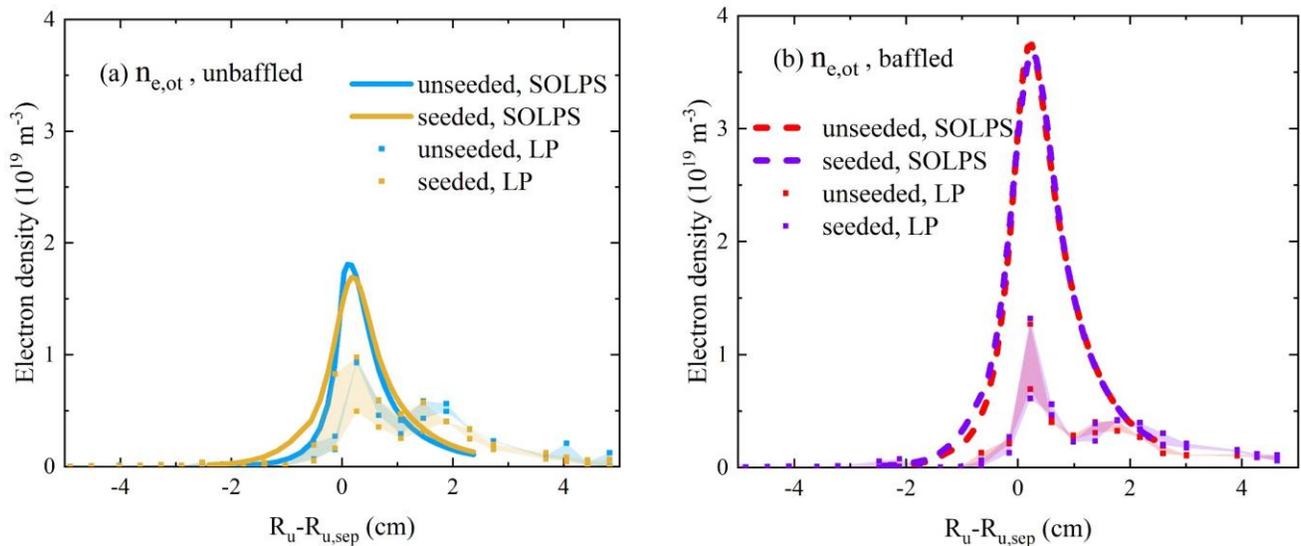

Figure 20. Comparison of LP-measured and simulated outer target electron density in the (a) unbaffled and (b) baffled TCV configurations. The nitrogen seeding rate in the seeded cases is $5.0 \times 10^{20}$ atom s$^{-1}$. Shaded areas represent the upper and lower limit of the LP measurements for the four TCV discharges shown in section section 4.1. Each point of the measurement corresponds to the density and $R_u - R_{u,sep}$ of one particular probe averaged over a time interval of 5 ms.

Target heat fluxes, $q_{ot}$, are measured using TCV's IR thermography diagnostic. Here the heat flux due to neutrals and radiation is included in the simulation results. The observed decrease of $q_{ot}$ with both baffles and is consistent with SOLPS simulations, Figure 21(a, b). The SOLPS predictions are comparable with IR measurement without baffles and seeding, but underestimate the heat flux for the other three cases. In addition, SOLPS predicts broader target heat flux profiles, in particular with baffles.

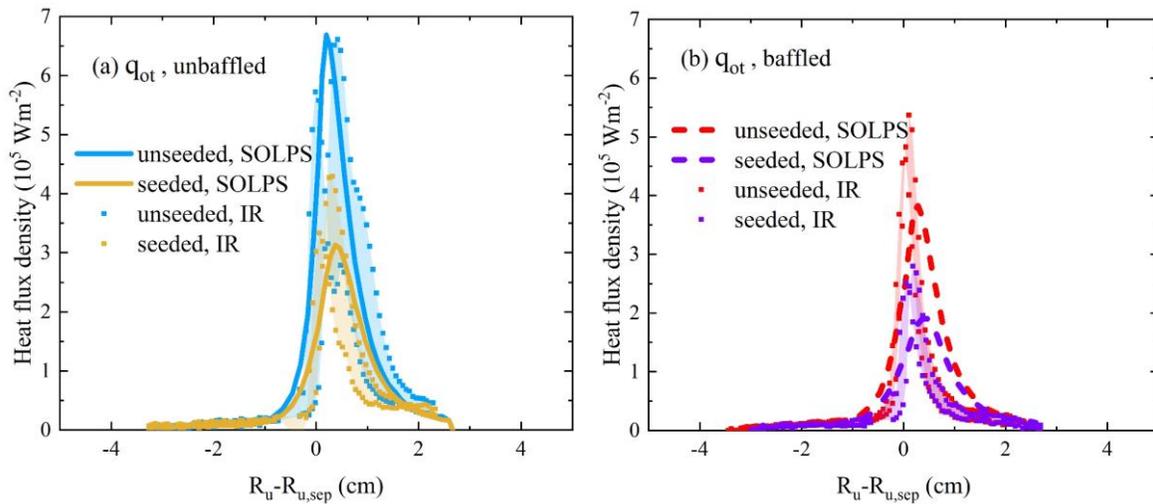

Figure 21. Comparison of IR-measured and simulated outer target plasma heat fluxes in the (a) unbaffled and (b) baffled TCV configurations. The nitrogen seeding rate in the seeded cases is $5.0 \times 10^{20}$ atom s$^{-1}$. Shaded areas represent the upper and lower limit of the IR measurements for the four TCV discharges shown in section 4.1. Each point of the measurement corresponds to the heat flux and $R_u - R_{u,sep}$ of one particular probe averaged over a time interval of 5 ms.

### 4.3. Comparison of neutral pressure

The detachment is primarily achieved by transferring momentum from the plasma to neutrals. The behavior of neutrals significantly influences the SOL properties. Baratron pressure gauges are installed in TCV to monitor the neutral pressure in divertor. Since the baratron gauges cannot operate under high magnetic field, they are located outside the toroidal field coils at the end of long extension tubes. The neutral conductance of the tubes introduces a nonnegligible time delay. Collisions of neutrals with the tube lead to a pressure drop along the tube, which must be taken into account in the comparison with SOLPS. The employed model was originally proposed for interpreting neutral dynamics in Alcator C-Mod [51] and is also applied to TCV modeling [27]

Baratron gauge measurements and SOLPS simulations, both, show a significant increase of divertor neutral pressure by baffles, while the divertor neutral pressure is independent of seeding, Figure 22(a-b)[4]. SOLPS overestimates the neutral pressure by a factor of approximately two without baffles and by a factor of 3-4 with baffles, indicating that the increase of divertor neutral pressure by baffles is overestimated by SOLPS. This overestimation of neutral pressure by simulation is consistent with abovementioned target profiles, i.e. simulation gives a denser and colder divertor than measured from diagnostics and similar

---

[4] The oscillation of baratron pressure measurements in the unbaffled discharges is thought to be caused by mechanical vibrations of the baratron gauge support

to previous findings without nitrogen seeding [38]. The compression is not measured as the outer board mid-plane pressure is below the baratron gauge measurement limit of approximately 2 mPa.

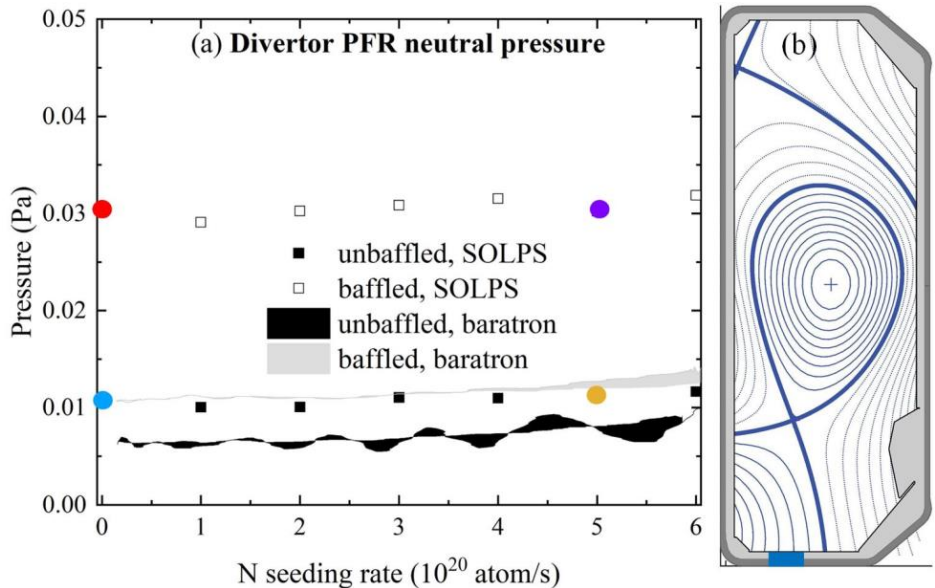

Figure 22 Comparison of measured and simulated neutral pressure in the unbaffled (solid) and baffled (dashed) TCV configurations. (a) Divertor PFR neutral pressure measured by the baratron gauge. (b) Locations of divertor baratron gauge in TCV. Shaded areas represent the upper and lower limit of the baratron gauges. Each point of the measurement corresponds to the pressure averaged over a time interval of 5 ms.

4.4 Comparison of radiated power

The radiated power distribution, measured with TCV's RADCAM bolometer system (BOLO) [52], and impurity line emission, measured with the Divertor Spectroscopy System (DSS) [53], are used to validate the observed effects of baffles and seeding on the impurity radiation.

In the analyzed discharges, the plasma radiation measurements along 120 chords of the BOLO diagnostic reveal the spatial radiation distribution. The cameras are located at the vessel top, in three lateral ports and at the bottom of the vessel, Figure 23(e). To compare the simulations with the measurement, the simulated radiated power density is integrated along the link-of-sight of each chord, similar to previous work [38]. Since SOLPS does not simulate the core plasma, predicted synthetic BOLO signals in chords facing the plasma core at the top, upper and middle lateral of the vessel (chords 1-60) are considerably lower than the measurement, particularly the middle lateral chords (yellow), Figure 23(a-d). Measured radiation in lower lateral and bottom chords (chords 61-120) facing the divertor are comparable with the SOLPS prediction except near the inner and outer strike points. SOLPS overpredicts the inner strike point radiation by a factor of two to three, for all three cases in Figure 23(a-

d) except the unbaffled, unseeded case. Measured radiation in the bottom and lower lateral chords increases with both baffles and seeding, consistent with the SOLPS predictions.

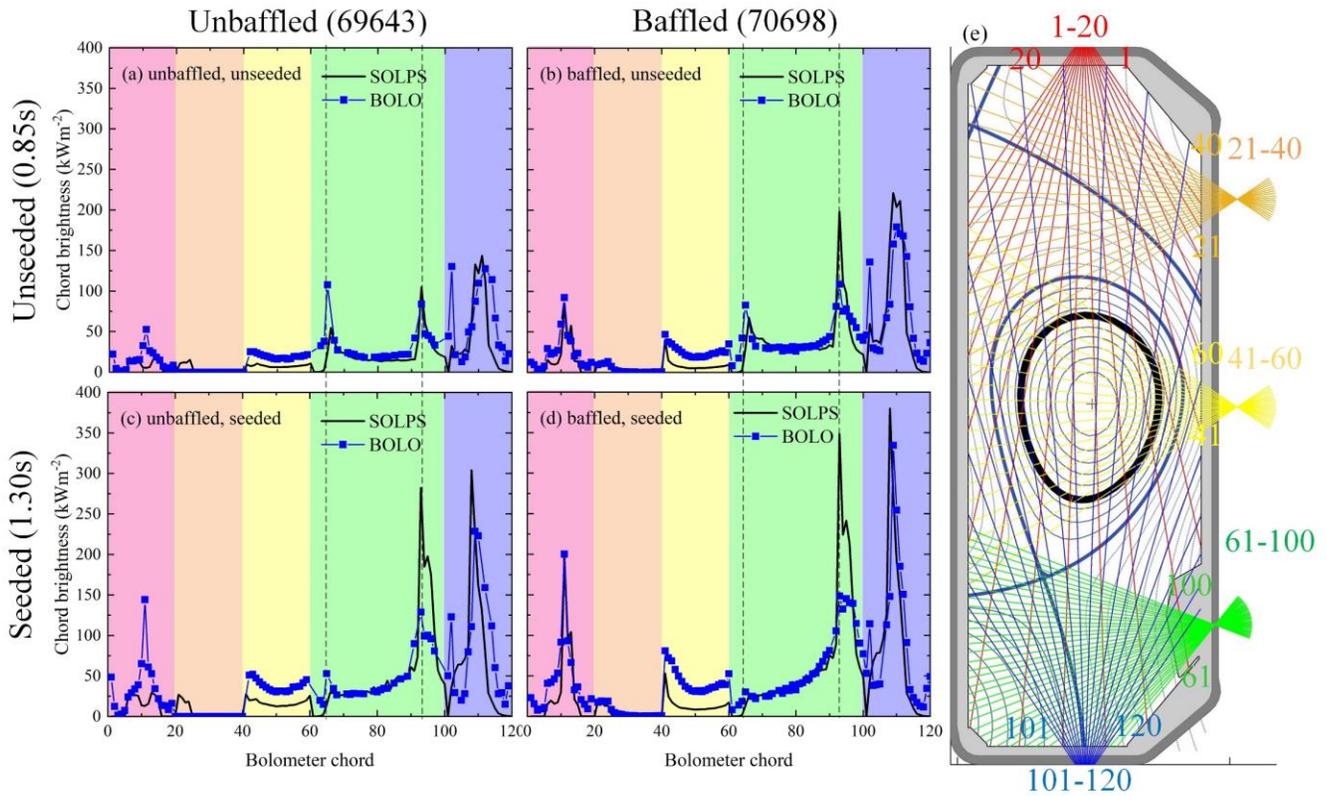

Figure 23. Comparison of measured and simulated bolometer chord brightness in the (a) unbaffled, unseeded case, (b) baffled, unseeded case, (c) unbaffled, seeded case and (d) baffled, seeded cases. The seeding rate in the seeded cases is $5.0 \times 10^{20}$ s$^{-1}$. (e) Bolometer coverage. Shaded zones in (a)-(d) correspond to BOLO cameras at top (red), upper lateral (orange), middle lateral (yellow), lower lateral (green), and bottom (blue) positions. Discharge 69643 misses measurements of the upper lateral camera, which results in a small underestimation of the radiated power. Inner and outer strike points are makred by dashed lines in the right and left of (a)-(d), respectively. The SOLPS core region boundary is marked by the solid black line in (e).

The 2D radiated power distribution can be obtained from the BOLO measurement by performing the tomographic inversions, which is used to calculate the divertor radiation by integrating the obtained radiation distribution over the divertor region (same definition as in section 3.2). The divertor radiation increases with baffles and seeding, Figure 24. The trends of measured divertor radiation with baffles and seeding are consistent with the SOLPS prediction, though the SOLPS simulations predict slightly stronger radiation than measured at high seeding levels.

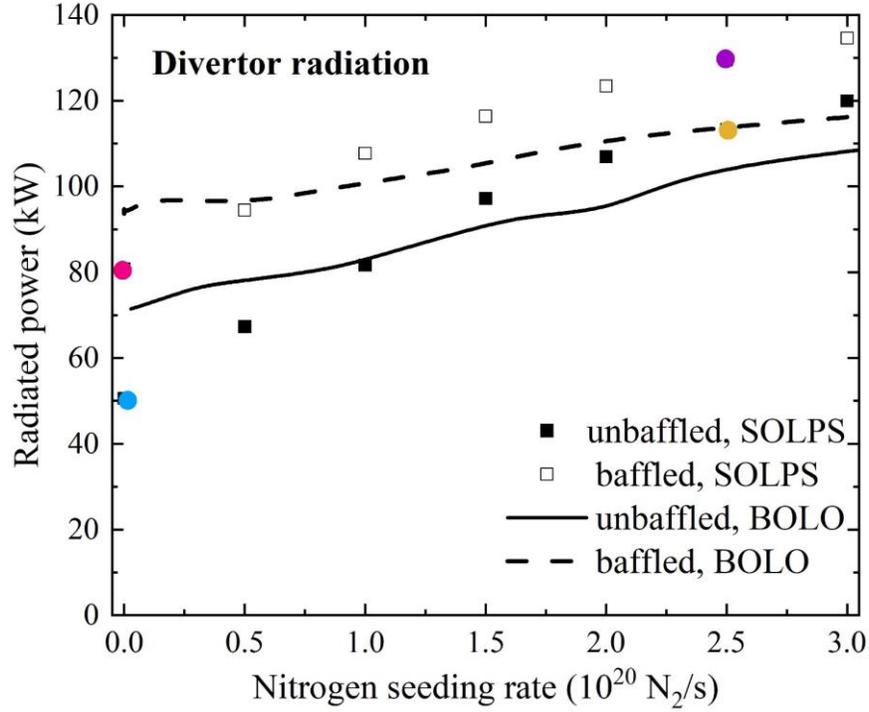

Figure 24. Comparison of the measured and simulated divertor radiation with varying nitrogen seeding rate. The SOLPS datapoints with various seeding rates are matched with the time axis according to Figure 17(a). The colored points correspond to the four simulation cases in Table II. Only measurement of one discharge is available for unbaffled and baffled divertor (69643 and 70698). The oscillations in the measurement are due to limited tomographic inversion frequency.

The DSS diagnostic measures the spectrally resolved emission along 30 chords intersecting the inner and outer divertor leg, Figure 26(g), [53]. By comparing the DSS measurements with the synthetic diagnostic signals in SOLPS, consistency of impurity emission lines can be surveyed to validate the changes of carbon and nitrogen emission with seeding and baffles. Here line-integrated intensity of each DSS chord for CII 426.8 nm line emitted by $C^+$ ions, and NII 399.5 nm line emitted by $N^+$ ions, are inspected. The line emissions feature two peaks in the first and last five chords, corresponding to the inner and outer targets. The outer strike point emission peaks disappear when the divertor is highly detached and the radiation fronts move away from the target, Figure 25(d-f).

The SOLPS predictions generally overpredict the CII emission by approximately a factor of two without baffles and four with baffles along the divertor leg, and more at the inner and outer strike points. In particular, the predicted increase of CII radiation with baffles is not observed. DSS measurements show that baffles increase the divertor CII emission only near the inner strike point, for both unseeded and seeded cases, Figure 25(a-d). SOLPS simulations show that baffles increase the divertor CII emission in most chords, except for the baffled, seeded case where the outer strike point emission is lower with baffles. Meanwhile, DSS measurements show that seeding decreases the divertor CII emission near the outer strike point, consistent with SOLPS simulations, Figure 25(a-d). The discrepancies are possibly related to the broader target heat flux profile which overpredicts the DSS line integral.

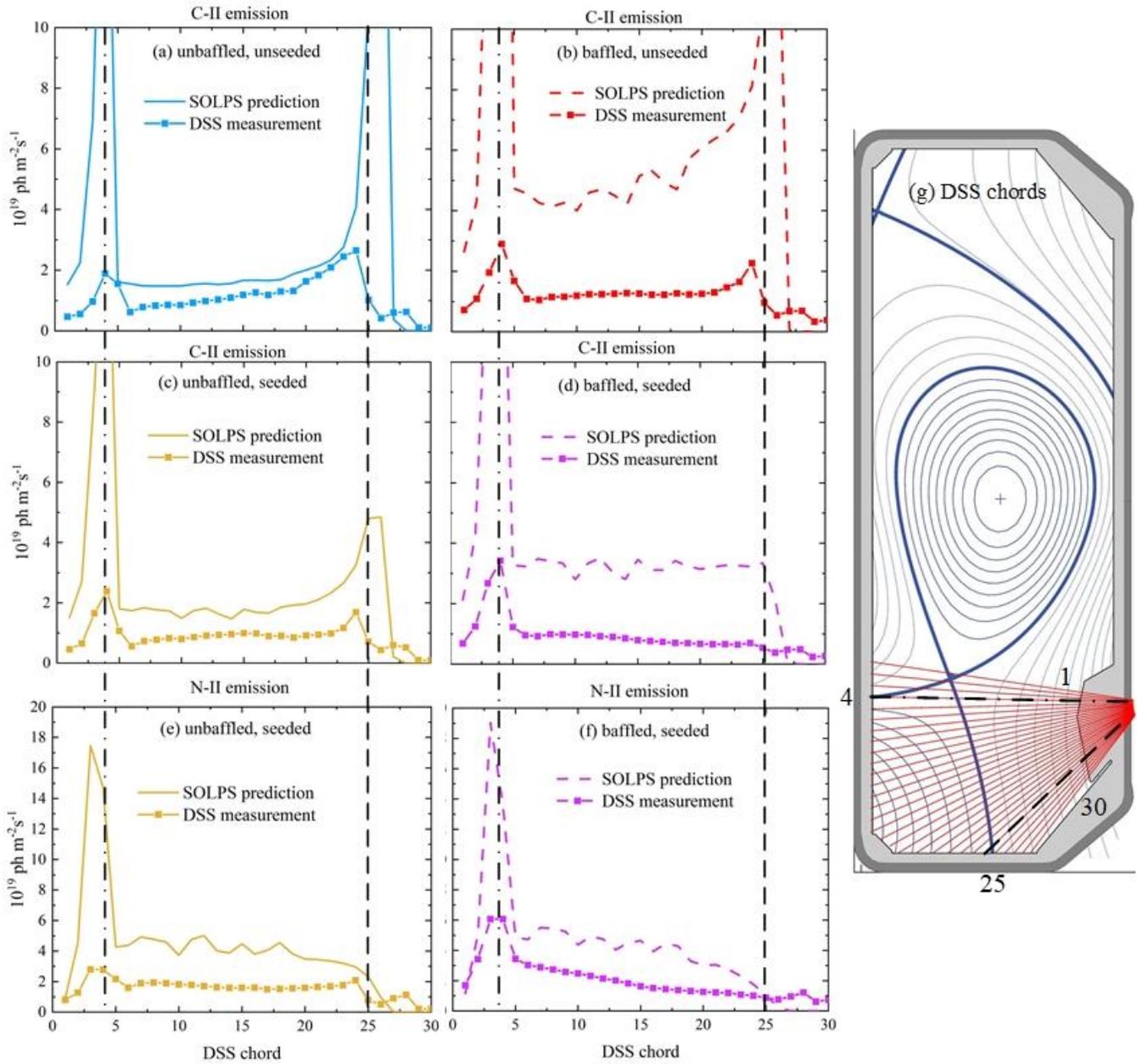

Figure 25. CII (426.8 nm) emission (a-d) and NII (398.5 nm) emission (e-f) comparison between DSS and synthetic diagnostic for (a) unbaffled, unseeded case, (b) baffled, unseeded case, (c) unbaffled, seeded case, (d) baffled, seeded case, (e) unbaffled, seeded case, and (f) baffled, seeded cases. (g) DSS chord number and line of sight. The seeding rate is $5.0 \times 10^{20}$ s$^{-1}$ in the seeded cases. Only measurement of one discharge is available for unbaffled and baffled divertor (69643 and 70698). The chords number 4 and 25 are marked by dashed-dotted lines and dahsed lines, respectively.

SOLPS simulations generally overestimate the NII emission by a factor of two to three along the divertor leg, and more at the inner strike point. DSS measurements confirm that baffles do not significantly increase the divertor NII emission, Figure 25(e-f). Seeding increases the NII emission, which is trivial and not shown here.

4.5. Summary of the simulation-experiment comparison

The main observations are:

1) The comparison verifies that baffles increase the target electron density and particle flux, while decreasing the target electron temperature and heat flux. In addition, seeding decreases the target electron temperature, particle flux and heat flux, but does not change the target electron density. SOLPS generally overpredicts the measured target density and, hence, particle fluxes.

2) The comparison verifies that baffles significantly increase the divertor neutral pressure, while the divertor neutral pressure is not affected by seeding. SOLPS generally overpredicts the measured neutral pressures.

3) The comparison verifies that the overall radiation increases with seeding, and increases with baffles with the increase more obvious without seeding. The comparison also verifies that baffles and seeding both increase the divertor radiation. It is found that baffles do not increase the CII emission as predicted by SOLPS, while the decrease of CII emission with seeding is verified. Baffles are found to have only a weak effect on the divertor NII emission, consistent with SOLPS-ITER predictions. SOLPS slightly overpredicts the radiated power at high seeding levels, and significantly overpredicts the line-integrated CII and NII emissions.

Potential reasons that can cause the discrepancies in the comparison as well as the possible measures to fix them in future works are briefly discussed. First, the nitrogen recycling coefficient, which directly affects the amount of nitrogen in the TCV chamber, is not well known and can even change within a discharge. This uncertainty could be at least partially mitigated by choosing another ordering parameter, such as the target temperature. Second, drifts are neglected and the chosen constant diffusivities neither optimized nor necessarily an appropriate model of the cross-field transport. Including drifts and tuning diffusivities to match measured profiles will be included in future work. Third, recent studies highlighted a stronger importance of plasma-molecule interaction with systematic errors in some of the included reaction cross sections in SOLPS [35]. Corrected cross-sections should decrease discrepancies. Finally, SOLPS-ITER simulates stationary plasma states, whereas the nitrogen seeding rate is dynamic and ramps up in TCV discharge. Experiments with varied nitrogen ramps should confirm when the plasma states can be considered quasi-static.

## 5. CONCLUSIONS

The interplay between baffling and nitrogen seeding in L-mode TCV divertor detachment was investigated with SOLPS-ITER simulations and tested with TCV experiments. The simulation-experiment comparison shows that outer target electron temperature and heat flux are significantly reduced by both baffles and seeding. Baffles increase the target particle flux and density, whereas seeding decreases the target particle flux. Contributions of baffles and seeding on target parameters are cumulative, and are explained by the two-point model.

The simulation-experiment comparison also verifies that baffles significantly increase the divertor neutral pressure and neutral compression. The roles of baffles on neutral distribution are explained by a schematic neutral transport model with SOLPS-ITER simulations, where baffles are shown to decrease the neutral conductance from the divertor to the main chamber.

SOLPS-ITER simulations further reveal the effects of baffles and seeding on the impurity distribution. SOLPS simulations predict that baffles can increase the divertor carbon density due to higher target particle flux, though the observed particle flux increase is weaker than predicted and no indications for an increased divertor carbon density seen. Baffles increase the nitrogen impurity retention and decrease the main chamber nitrogen density at high seeding levels. This is explained by the changes of main ion flow by baffles and the changes of ion temperature by baffles and seeding. The direct effects of baffles on nitrogen neutrals is significant for the divertor nitrogen density, but not the main chamber nitrogen density. An experimental verification would require main chamber nitrogen measurements, which is expected to be conducted in future works. The changes of impurity density distribution affect the radiation accordingly. The simulation-experiment comparison shows that baffles and seeding both increase the divertor radiation. Baffles do not considerably increase the CII emission and NII emission. Seeding decreases the divertor CII emission and increases the NII emission.

The first comparison of SOLPS-ITER simulations and TCV experiments with baffles and seeding shows qualitative consistency in trend, though the simulations predict a colder and denser divertor compared with experiments. The discrepancies are expected to be reduced in future works by, e.g. including drifts, refining choices of transport coefficient and nitrogen recycling coefficient in simulation, optimizing the strategy for simulation-experiment comparison, etc. The present work increases the confidence of using SOLPS-ITER simulations for the next TCV divertor upgrade.

ACKNOWLEDGMENTS

This work has been carried out within the framework of the EUROfusion Consortium, via the Euratom Research and Training Programme (Grant Agreement No. 101052200— EUROfusion) and funded by the Swiss State Secretariat for Education, Research and Innovation (SERI). Views and opinions expressed are however those of the author(s) only and do not necessarily reflect those of the European Union, the European Commission, or SERI. Neither the European Union nor the European Commission nor SERI can be held responsible for them. This work was s supported in part by the Swiss National Science Foundation.

DATA AVAILABILITY STATEMENT

The data that support the findings of this study are available from the corresponding author upon reasonable request.